\documentclass{artcl}

\hypersetup{
  pdftitle={Exact fermionic matrix elements from the {Bethe-ansatz} solution of the
  {Richardson--Gaudin} model},
  pdfauthor={Peter D. Johannsen, Valerii K. Kozin, Jelena Klinovaja, Daniel Loss},
  pdfkeywords={Richardson--Gaudin, Eigenvalue-based variables, Broken integrability},
  pdfsubject={
    We introduce an effective Hamiltonian description of mesoscopic superconducting systems
    based on the Bethe-ansatz solution of the Richardson--Gaudin model. Our central result
    is a compact determinant formula for the matrix elements of fermionic creation and
    annihilation operators, expressed entirely in terms of eigenvalue-based variables. This
    formula allows single-electron tunneling terms to be represented directly in the
    eigenbasis of the Richardson--Gaudin model, extending the eigenvalue-based framework
    beyond overlaps and previously known form factors. As a benchmark, we apply the method
    to a superconducting Anderson impurity model comprising a quantum dot coupled to an
    ultrasmall superconducting grain. The resulting low-energy description, formulated
    naturally in the canonical ensemble, captures even-odd effects, subgap excitations, and
    quantum-dot physics. The same framework gives access to observables not previously
    computed for such grain-dot systems in the canonical ensemble, such as the
    fixed-particle-number optical conductivity and flux-tunable spectra.
  },
}

\usepackage{xcolor}
\colorlet{BLUE}{blue}

\newcommand{\dl}[1]{{\color{red}\ [DL: #1]}}
\newcommand{\hrg}{\ensuremath{\hat H_{\rm RG}}\xspace}
\newcommand{\hcharge}{\ensuremath{\hat H_{\rm charg.}}\xspace}
\newcommand{\hhop}{\ensuremath{\hat H_{\rm tunn.}}\xspace}
\newcommand{\hhopphase}{\ensuremath{\hat H_{{\rm tunn.},A(t)}}\xspace}
\newcommand{\V}{\ensuremath{\mathcal V}\xspace}
\newcommand{\curr}{\ensuremath{\skew4\hat j}\xspace}
\newcommand{\hhopflux}{\ensuremath{\hat H_{{\rm tunn.},\Phi}}\xspace}
\newcommand{\htun}{\ensuremath{\hat H_{\lambda}}\xspace}
\newcommand{\hcond}{\ensuremath{\hat H_{\rm cond.}(t)}\xspace}
\newcommand{\hrga}{\ensuremath{\hat H_{\rm RGA}}\xspace}
\newcommand{\hrgaphi}{\ensuremath{\hat H_{{\rm RGA}, \Phi}}\xspace}

\newcommand{\qm}{\ensuremath{\hat Q_m}\xspace}
\newcommand{\ebv}{\ensuremath{\Lambda}}
\newcommand{\cop}{\ensuremath{{\hat{c}}}}
\renewcommand{\d}{\ensuremath{{\hat{d}}}}
\newcommand{\ydim}{\mathfrak D}

\def\andersonsubgap{a}
\def\andersonphasetransition{b}
\def\andersonweak{c}
\def\andersonstrong{d}

\def\fluxepr{a}
\def\fluxcpr{b}
\def\fluxdome{c}
\def\fluxweak{d}
\def\fluxstrong{e}
\def\fluxdepiction{f}

\def\condboth{a}
\def\condodd{b}
\def\condeven{c}
\def\condfig{d}
\def\condgap{e}
\def\condsubgap{f}

\makeatletter
\def\env@cases{%
  \left\lbrace
  \def\arraystretch{1.2}%
  \array{@{}l@{\hspace{2ex}}l@{}}%
}
\makeatother
\begin{document}


\title{Exact fermionic matrix elements from the {Bethe-ansatz} solution of the
{Richardson--Gaudin} model}

\def\unibas{Department of Physics, University of Basel, Klingelbergstrasse 82, 4056
Basel, Switzerland}

\author{Peter D. Johannsen}
\affiliation{\unibas}

\author{Valerii K. Kozin}
\affiliation{\unibas}

\author{Jelena Klinovaja}
\affiliation{\unibas}

\author{Daniel Loss}
\affiliation{\unibas}
\affiliation{Physics Department, King Fahd University of Petroleum and Minerals, 31261,
Dhahran, Saudi Arabia}
\affiliation{Center for Advanced Quantum Computing, KFUPM, Dhahran, Saudi Arabia}
\affiliation{RDIA Chair in Quantum Computing}

\date{\today}


\begin{abstract}
  We introduce an effective Hamiltonian description of mesoscopic superconducting systems
  based on the Bethe-ansatz solution of the Richardson--Gaudin model. Our central result
  is a compact determinant formula for the matrix elements of fermionic creation and
  annihilation operators, expressed entirely in terms of eigenvalue-based variables. This
  formula allows single-electron tunneling terms to be represented directly in the
  eigenbasis of the Richardson--Gaudin model, extending the eigenvalue-based framework
  beyond overlaps and previously known form factors. As a benchmark, we apply the method
  to a superconducting Anderson impurity model comprising a quantum dot coupled to an
  ultrasmall superconducting grain. The resulting low-energy description, formulated
  naturally in the canonical ensemble, captures even-odd effects, subgap excitations, and
  quantum-dot physics. The same framework gives access to observables not previously
  computed for such grain-dot systems in the canonical ensemble, such as the
  fixed-particle-number optical conductivity and flux-tunable spectra.
\end{abstract}
\maketitle


\section{Introduction}
Superconductivity in mesoscopic devices exhibits a rich range of physical phenomena, such
as the competition between superconducting pairing and local-moment formation, fermionic
parity effects, and Andreev bound states that carry supercurrent~\cite{vondelft2001,
  sierra2000a, pavesic2021, pavesic2022, ralph1995, ralph1997,
  black1996, jorgensen2007, vandam2006, zonda2016, lim2008, lee2014, andreev1964,
andreev1966, meng2009, maurand2012}. Such hybrid mesoscopic devices containing
superconductors have attracted particular interest in recent years, owing to their
potential applicability to quantum computation~\cite{park2017, hays2021}.

A particularly important regime is that of ultrasmall superconducting grains, where the
superconducting gap is not parametrically large compared to the single-particle level
spacing, and the grand-canonical mean-field formulation of Bardeen--Cooper--Schrieffer
(BCS) theory becomes inadequate~\cite{vondelft2001}. Additionally, when charging effects
are appreciable, the canonical ensemble is the appropriate framework for describing these
devices~\cite{pavesic2021, pavesic2022}. In this regime, the reduced BCS Hamiltonian and
its Bethe-ansatz solution offer a natural starting point, as they intrinsically describe
the superconducting subsystem at fixed particle number.

The same physics is also relevant to granular superconductors such as granular aluminium,
which consists of networks of interconnected nanoscale superconducting
grains~\cite{grunhaupt2019, rieger2023}. This makes such systems a natural setting for
number-conserving methods.

\begin{figure}
  \centering
  \includegraphics[width=0.8\linewidth]{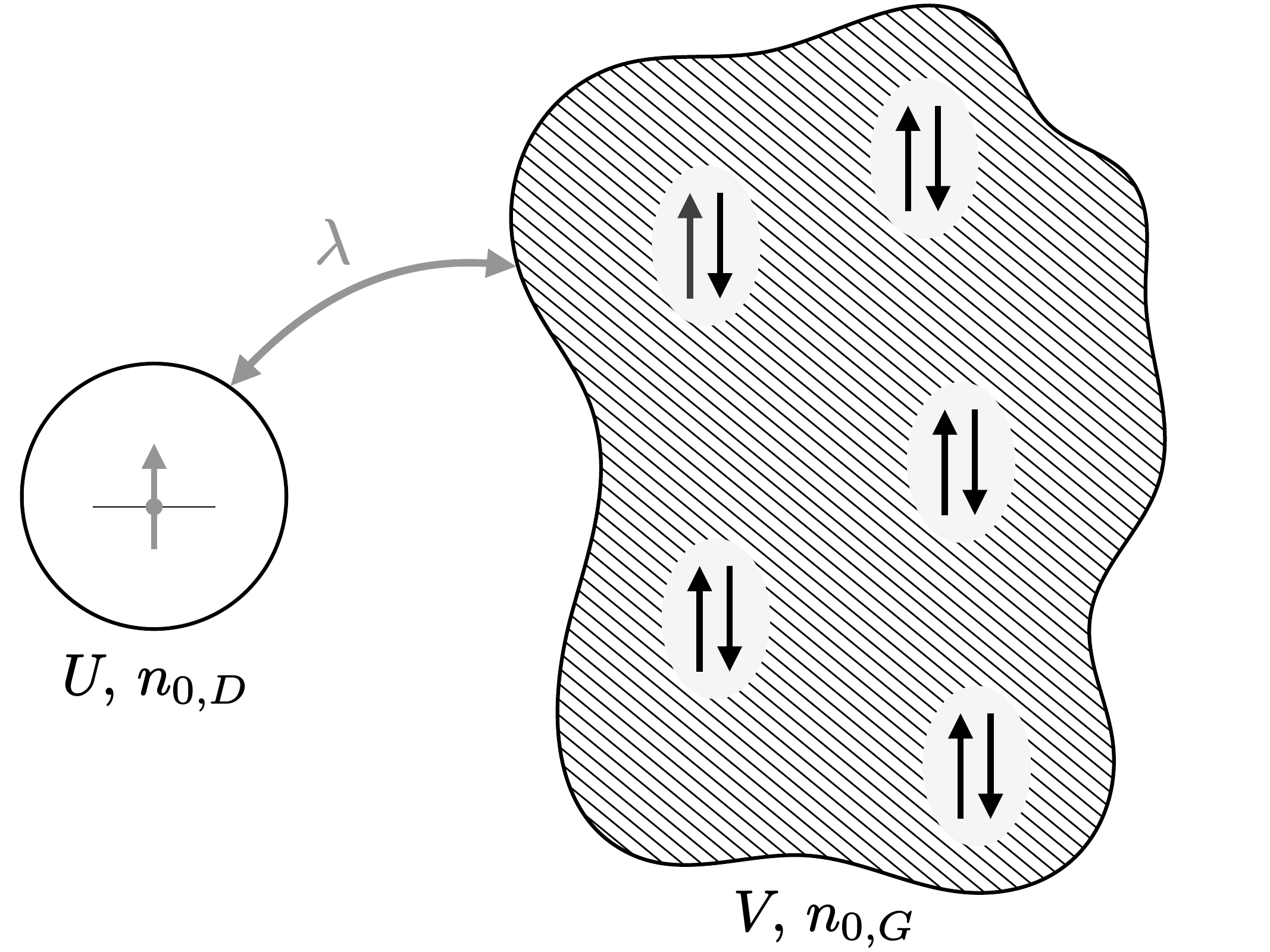}
  \caption{An ultrasmall superconducting grain (hatched fill) coupled to a single dot
  level.}\label{fig:setup}
\end{figure}

However, typical mesoscopic superconducting devices do not consist of isolated grains but
are hybrid structures in which single-electron tunneling plays an important role. In this
work, we focus on grain-dot systems, shown schematically in Fig.~\ref{fig:setup}, in which
an ultrasmall superconducting grain is coupled to a single quantum-dot level. In such
systems, the integrability of the reduced BCS Hamiltonian generally cannot be exploited
because single-electron tunneling breaks integrability~\cite{pavesic2021}. We
nevertheless leverage the integrability of the BCS Hamiltonian by projecting the full
Hamiltonian onto a truncated basis of exact Richardson--Gaudin eigenstates. Thus, this
work serves as a proof of principle for bridging the gap between studying integrable
models for their mathematical structure and describing experimentally relevant mesoscopic
devices. To achieve this, we derive an eigenvalue-based determinant expression for matrix
elements of individual fermionic creation and annihilation operators, which connects
sectors differing by one electron and provides direct access to single-electron tunneling.

The method builds on existing results that apply only to inner products and selected form
factors~\cite{faribault2012, faribault2022, claeys2017, claeys2015a, claeys2015,
johnson2025}. For instance, Ref.~\cite{johnson2025} provides eigenvalue-based
constructions for number-conserving one- and two-body operators at nonzero seniority. For
number-changing single-fermion operators, Ref.~\cite{gorohovsky2014} expresses matrix
elements through Slavnov overlaps and derives analytical expressions for coarse-grained
squared matrix elements in the thermodynamic limit, generalizing the BCS coherence factor
to two-arc Richardson states. Our contribution is instead an exact finite-size,
eigenvalue-based determinant expression for such matrix elements between arbitrary
states, which was, to our knowledge, not previously available. A related RG-eigenbasis
approach to weak integrability breaking was presented in Ref.~\cite{claeys2017b}, which
studies a Richardson--Gaudin-like system in which nonuniform attractive interactions
instead break integrability.

Charge-conserving density-matrix renormalization-group (DMRG) methods treat such
grain-impurity systems directly and reach large sizes~\cite{dukelsky1999, pavesic2021,
pavesic2022}. Our approach is complementary: it uses exact grain eigenstates and
closed-form fermionic matrix elements within a small retained basis, making projected spectra,
response functions, and dense parameter sweeps inexpensive to calculate.

The ability to compute fermionic matrix elements between Richardson--Gaudin eigenstates
may, moreover, enable applications to nuclear structure observables, such as
single-particle transfer amplitudes and spectroscopic factors.

This work is structured as follows: Sec.~\ref{sec:model} introduces the reduced BCS
Hamiltonian and its Bethe-ansatz solution, and presents the conserved charges and the
description of the model in terms of eigenvalue-based variables (EBVs). The section
continues with a determinant expression that generalizes the results of
Refs.~\cite{faribault2012, claeys2017} to states of arbitrary seniority. We next present
our central result --- a similar determinant expression for the matrix elements of
fermionic creation and annihilation operators --- together with a discussion of the
state-selection procedure used in the low-energy projection. This methodology is then
applied in Secs.~\ref{sec:anderson} and~\ref{sec:flux}, where we discuss the
superconducting Anderson impurity model without and with flux tunability, respectively.


\section{Model and Method}\label{sec:model}
We study the rational Richardson--Gaudin (RG) pairing model, defined by the reduced BCS
Hamiltonian~\cite{richardson1963, richardson1964, richardson1966, gaudin1976,
dukelsky2004, debaerdemacker2014, bardeen1957}:
\begin{align}
  \hrg = \sum_{m = 1}^N 2\epsilon_m \left(\hat S_m^z + \frac12 \right) - g
  \sum_{m,\ell=1}^N \hat S^+_m \hat S^-_\ell, \label{eq:hrg}
\end{align}
where the pseudospin operators are defined as $\hat S_m^z =
(\sum_{\sigma=\uparrow,\downarrow} \cop^\dagger_{m \sigma} \cop_{m \sigma} - 1)/2$, $\hat
S^+_m = \cop^\dagger_{m \uparrow} \cop_{m \downarrow}^\dagger$ and $\hat S^-_m =
\cop_{m\downarrow} \cop_{m\uparrow}$. Here, $\cop_{m \sigma}$ annihilates an electron in
level $m$ with spin $\sigma$ in the superconducting grain, $N$ denotes the number of
spin-degenerate single-particle energy levels, and $g>0$ is the attractive interaction
strength. Throughout this work, $\epsilon_m$ denotes a single-particle energy; we set
$\epsilon_{m} = md$, where $d$ denotes the single-particle level spacing\footnote{Thus,
  $2 \epsilon_m$ denotes the noninteracting energy of a pair in level $m$, and distinct
  fixed-pair-number energies lie on a $2d$ grid at $g = 0$ (the many-body spectrum is
degenerate, not a uniformly spaced ladder).}. Our methods apply to arbitrary
nondegenerate single-particle energy levels; however, we adopt the standard choice of
equidistant single-particle energies, commonly referred to as the `picket-fence model'.

As a reference scale, we use the standard half-filled, symmetric-band BCS
expression~\cite{roman2002} for the single-particle gap $\Delta = \Omega / [2
\sinh(d/g)]$, where $\Omega = Nd$ is the single-particle bandwidth\footnote{This is the
continuum-cutoff convention; the exact span of the levels is $(N-1)d$.} at $g = 0$.
Although the systems studied in this work contain $N = 30$ levels, $\Delta$ serves as an
appropriate reference energy scale. The low-energy projection introduced below is
controlled by the ratio of the coupling between retained and discarded sectors to their
energy separation; in the parameter regime studied here this ratio is small, as assessed
in App.~\ref{app:convergence}.

The excitations from the ground state of \hrg that we study in this work fall into two
categories: (1) single-particle excitations, where a single electron is either added to
or removed from the ground state; and (2) particle-number-conserving excitations, where
the number of electrons in the excited state equals the number of electrons in the ground
state. In this work we study both types: the optical conductivity probes neutral
excitations, while the particle-addition and -removal spectra probe single-particle excitations.

\subsection{Bethe-ansatz solution}
The reduced BCS Hamiltonian is integrable, despite being quartic in the fermionic
operators. In fact, all eigenstates can be expressed explicitly in terms of the Bethe
ansatz, which is parametrized by $M_\alpha$ rapidities $u_{\nu}^\alpha \in \mathbb
C$~\cite{richardson1963}:
\begin{align}
  \ket{\Psi_{\alpha}} = A_\alpha \prod_{(b, \sigma_b) \in \C_{\alpha}} \cop_{b
  \sigma_b}^\dagger \left[ \prod_{\nu = 1}^{M_\alpha}{\frac g2\sum_{m \in
  \nB_{\alpha}}}\; \frac{\hat S_m^+}{\epsilon_m - u_\nu^{\alpha}} \right] \ket\vac,
  \label{eq:bethe_ansatz}
\end{align}
where $\ket\vac$ is the fermionic vacuum. We define $\C_{\alpha}$ as the set of
blocked-level tuples, whose elements $(b, \sigma_b)$ specify the index of the blocked
level $b$ and the spin $\sigma_b$ of the unpaired fermion occupying that level. Products
over $\C_\alpha$ are ordered by increasing blocked-level index. The set of blocked levels
is defined as $\B_\alpha = \{b:(b,\cdot) \in \mathcal C_\alpha\}$, and $\nB_\alpha$
denotes its complement (i.e., the set of unblocked levels). The number of pairs
$M_\alpha$, together with the set of blocked levels, determines the number of electrons
in the state, $n_{G, \alpha} = \langle \Psi_\alpha \vert \hat n_G \vert \Psi_\alpha
\rangle = 2 M_\alpha + \vert \B_\alpha \vert$. Here, $\hat n_G = \sum_{m = 1}^N
\sum_{\sigma = \uparrow,\downarrow} \cop^\dagger_{m\sigma} \cop_{m\sigma}$ is the
grain-number operator. Moreover, the \textit{seniority} of the state is defined as the
number of unpaired electrons, which equals $\vert \C_{\alpha} \vert = \vert \B_\alpha
\vert$. The normalization factor $A_\alpha$, chosen real and positive, can be calculated
using the Gaudin determinant; however, we will give the alternative form in terms of the
eigenvalue-based variables below~\cite{claeys2017}.

For $\ket{\Psi_{\alpha}}$ to be an eigenstate of \hrg, the rapidities must satisfy the
\textit{Richardson equations}~\cite{richardson1963}:
\begin{align}
  \frac2g - {\sum_{m \in \nB_\alpha}} \frac{1}{\epsilon_m - u_\nu^{\alpha}} + \sum_{\mu
  \neq \nu} \frac{2}{u_\mu^{\alpha} - u_\nu^\alpha} = 0,\label{eq:richardson}
\end{align}
for all $1 \leq \nu \leq M_\alpha$. Within a fixed blocked-level sector, the rapidities
completely parametrize the state $\ket{\Psi_\alpha}$. In particular, they determine its
energy, $E_\alpha = 2\sum_{\nu=1}^{M_\alpha} u_\nu^\alpha + \sum_{b \in \B_\alpha}
\epsilon_b$\footnote{Although the rapidities are generally complex, the energy is
  guaranteed to be real because rapidities are either real or come in complex-conjugate
pairs.}, as well as its inner products and matrix elements, which can be expressed in
terms of Slavnov determinants; see Refs.~\cite{claeys2017, korepin1993, zhou2002, gorohovsky2014}.

\begin{figure}[tbp]
  \centering
  \includegraphics[width=\linewidth]{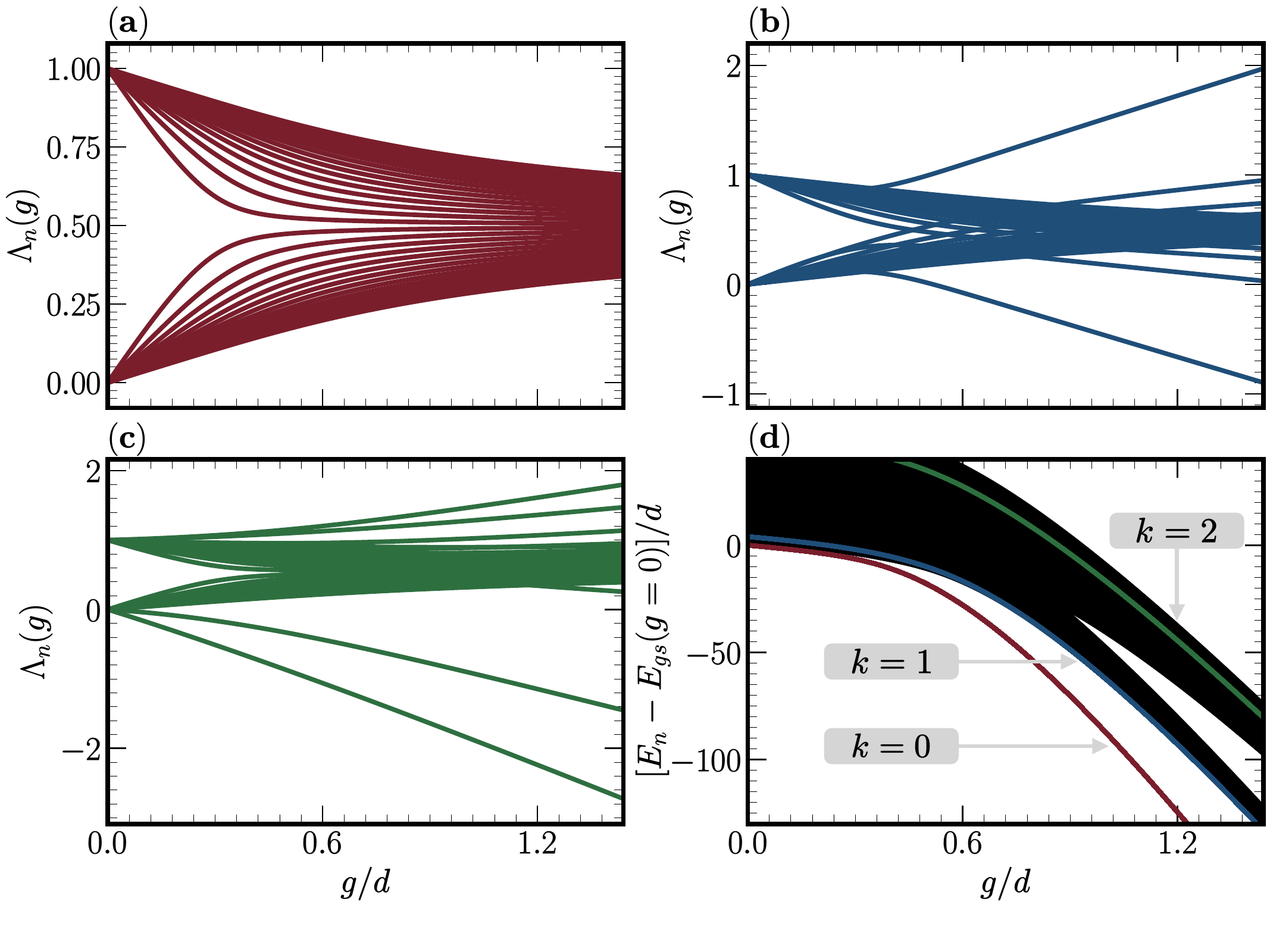}
  \caption{
    Eigenvalue-based variables (EBVs) for the ground state (a), and for states in the
    first and second excited bands [(b) and (c), respectively]. The energies of all
    seniority-zero states with excitation degree $k \leq 2$ [$k$ labels the large-$g$
    excitation bands, with $k=0$ for the ground state; see Sec.~\ref{sec:state-selection}] are
    shown in panel (d). The ground-state energy at $g=0$ --- denoted $E_{gs}(g = 0)$ ---
    is subtracted from all energies shown in (d). The system contains $N=30$ equidistant
    single-particle levels and $n_G = 30$ electrons. The red, blue, and green lines in
    (d) correspond to the energies of the states whose EBVs are displayed in panels (a),
    (b), and (c).
  }\label{fig:ebvs_and_energies}
\end{figure}
\subsection{Conserved charges \& eigenvalue-based variables}
An alternative approach to diagonalizing Eq.~\eqref{eq:hrg} uses $N$ \textit{conserved
charges} \qm. For unblocked levels, one can define~\cite{cambiaggio1997}:
\begin{align}
  \qm^\alpha = \left( \hat S_m^z + \frac12 \right) - g {\sum_{\ell \in \nB_{\alpha}
  \setminus \{m\}}} Z_{m\ell} \left( \hat {\bm S}_m \cdot \hat{\bm S}_\ell - \frac14
  \right),\label{eq:conserved_charges}
\end{align}
where $Z_{m\ell} \equiv (\epsilon_m - \epsilon_\ell)^{-1}$. For blocked levels, however,
$\hat Q_{b}^\alpha = 1/2$ for all $g$. The conserved charges commute mutually,
$[\qm^\alpha, \hat Q_n^\alpha] = 0$, and with the Hamiltonian, $[\qm^\alpha, \hrg] = 0$
\footnote{
  Removing the scalar terms from the definition of $\qm^\alpha$ gives the conserved
  charges the same form for \textit{all} blocked-level configurations, allowing the
  omission of the $\alpha$ superscript. However, we include the scalar terms so that the
  conserved charges annihilate the vacuum, in accordance with existing literature, apart from
  the factor of $g/2$ discussed below; see
  Refs.~\cite{faribault2012,faribault2022,claeys2017,claeys2015a,claeys2015,elaraby2012,johnson2025}.
}.

The eigenstates of $\hrg$ can also be chosen as eigenstates of the conserved charges,
$\qm^\alpha \ket{\Psi_\alpha} = \ebv_m^\alpha \ket{\Psi_\alpha}$, where the eigenvalues
$\ebv_m^\alpha$ are referred to as \textit{eigenvalue-based variables} (EBVs). Crucially,
the quadratic identities that $\qm^\alpha$ satisfy imply that the EBVs (for unblocked
levels) satisfy the following set of coupled equations:
\begin{align}
  F_m(\bm \Lambda^\alpha) = \ebv_{m}^\alpha\;(\ebv_m^\alpha - 1) - \frac g2 \!\!
  \sum_{\ell \in \nB_\alpha \setminus \{m\}} \!\! Z_{\ell m}(\ebv_{\ell}^\alpha -
  \ebv_{m}^\alpha) = 0. \label{eq:ebv}
\end{align}
Additionally, $\ebv_{m}^\alpha \in \mathbb R$ and $\sum_{\ell \in \nB_\alpha}
\ebv_{\ell}^\alpha = M_\alpha$. For blocked levels, $\ebv_{b}^\alpha = 1/2$ for all $g$.
As discussed in Ref.~\cite{elaraby2012}, the EBVs of unblocked levels ($m \in
\nB_\alpha$) can be expressed in terms of the rapidities, $\ebv_{m}^\alpha = \frac g2
\sum_{\nu=1}^{M_\alpha} (\epsilon_{m} - u_{\nu}^{\alpha})^{-1}$. Importantly,
Eq.~\eqref{eq:ebv} is numerically much better behaved than Eq.~\eqref{eq:richardson},
because it does not contain the rapidity singularities present in the Richardson
equations~\cite{elaraby2012}. This computational advantage is essential to the numerical
feasibility of our approach.

The EBVs used in this work are scaled by a factor of $g/2$ relative to those used in
Refs.~\cite{faribault2012,faribault2022,claeys2017,claeys2015a,claeys2015}, so that they
directly represent the eigenvalues of the conserved charges. At $g = 0$ the values of
$\ebv_m^\alpha - 1/2$ correspond exactly to the $z$ component of the Anderson
pseudospin~\cite{anderson1958} of the level $m$: $+1/2$ for doubly occupied levels, and
$-1/2$ for unoccupied levels. This is discussed further in Sec.~\ref{sec:state-selection}.

The EBVs can also be used to calculate the energy of each eigenstate
$\ket{\Psi_{\alpha}}$~\cite{faribault2022}:
\begin{align}
  E_\alpha = \sum_{m=1}^N 2 \epsilon_m \ebv_m^\alpha - g M_\alpha (N_\alpha^{\rm eff} -
  M_\alpha + 1),
\end{align}
where $M_{\alpha}$ denotes the number of pairs in the state, and $N^{{\rm eff}}_{\alpha}
\equiv \vert \nB_{\alpha}\vert$ denotes the number of unblocked levels. The $m$-sum above
includes blocked levels, for which $\ebv_{m}^\alpha = 1/2$. The term $g M_\alpha
(N_\alpha^{\rm eff} - M_\alpha + 1)$ is important when comparing states with different
seniority, as required for the effective Hamiltonians in the following sections. This
additional scalar term is often omitted in the literature, as it is irrelevant when
studying equal-seniority states with the same particle number. Numerically evaluated EBVs
for low-energy seniority-zero states are shown in Fig.~\ref{fig:ebvs_and_energies}(a--c),
with corresponding energies shown in Fig.~\ref{fig:ebvs_and_energies}(d).

Moreover, as discussed in Ref.~\cite{claeys2017}, the wave functions can be described
entirely in terms of the EBVs:
\begin{align}
  \ket{\Psi_\alpha} = A_\alpha \prod_{(b,\sigma_b) \in \C_\alpha} \cop^\dagger_{b
  \sigma_b} \sum_{P \in \mathcal P_\alpha} \det[{J_{\alpha,P}'}]\prod_{m \in P} \hat
  S_m^+ \ket\vac, \label{eq:bethe_ebv}
\end{align}
where $\mathcal P_\alpha$ is the set of distinct subsets of $\nB_\alpha$ with cardinality
$M_\alpha$, and $J'_{\alpha,P} \in \mathrm{Mat}(M_\alpha, \mathbb R)$ is defined as
\begin{align}
  (J'_{\alpha, P})_{mn} =
  \begin{cases}
    \ebv_{P_m}^\alpha - \frac g2 \sum_{\ell \neq m}Z_{P_mP_\ell}, & \text{ for } m = n,\\
    \frac g2 Z_{P_m P_n}, & \text{ for } m \neq n,
  \end{cases}\label{eq:Jprime}
\end{align}
where $P_m$ denotes the $m$th element of $P \in \mathcal P_\alpha$ (the expression does
not depend on the ordering of the elements within $P$). The existence of this
representation of the Bethe-ansatz wave function implies that all matrix elements can be
parametrized in terms of $\ebv_m^\alpha$, $\epsilon_m$, and $g$ alone.

For instance, inner products between Bethe-ansatz states can be expressed as
\begin{align}
  \bra {\Psi_\alpha} \ket{\Psi_\beta} = (-1)^{N^{{\rm eff}} + M}A_\alpha A_\beta
  \Sigma_{\alpha\beta}\det J_{\alpha\beta},\label{eq:innerproduct}
\end{align}
where, as shown in Refs.~\cite{faribault2012, faribault2022, claeys2015a, claeys2017},
for $m,n\in \nB_\alpha$
\footnote{
  $\C_{\alpha} = \C_{\beta}$; otherwise the inner product
  is zero; see App.~\ref{app:matrix_element_c}. Furthermore, $N_\alpha^{\rm eff} =
  N_\beta^{\rm eff} = N^{\rm eff}$ and $M_\alpha = M_\beta = M$.
}, $J_{\alpha\beta} \in \mathrm{Mat}(N^{\rm eff},\mathbb R)$ is defined as
\footnote{
  In Refs.~\cite{claeys2015a, claeys2017} the sign factor is $(-1)^M$. Our additional
  factor of $(-1)^{N^{\rm eff}}$ arises upon transforming from their conventions to ours:
  $\ebv \to g\ebv/2$, which rescales the EBVs, and $g \to -g$, reflecting that they take
  $g<0$ for attractive interactions whereas we use $g>0$.
}:
\begin{align}
  (J_{\alpha \beta})_{mn} =
  \begin{cases}
    \ebv_m^\alpha + \ebv_m^\beta - 1 + \frac g2 \displaystyle \!\!\!\sum_{\ell \in
    \nB_\alpha \setminus \{m\} }\!\!\! Z_{\ell m} & \text{for } m = n; \\
    \frac g2 Z_{mn} & \text{for } m \neq n,
  \end{cases}\label{eq:J}
\end{align}
and where
\begin{align}
  \Sigma_{\alpha\beta} = \bra\vac \left[\prod_{(n, \sigma_n) \in \C_\alpha} \cop_{n
  \sigma_n}^\dagger\right]^\dagger \prod_{(m, \sigma_m) \in \C_\beta} \cop^\dagger_{m
  \sigma_m} \ket\vac;
\end{align}
thus $\Sigma_{\alpha \alpha} = \Sigma_{\beta \beta} = 1$, and $\Sigma_{\alpha\beta} \in
\{0, 1\}$. The sign factor $\Sigma_{\alpha\beta}$ is the scalar part of the
normal-ordered blocked-level operator product; it can be interpreted as a blocked-level
vacuum expectation value, since for $\B_\alpha = \B_{\beta}$ the Hilbert space factorizes
into blocked- and unblocked-level subspaces. There exists a simple relation between the
matrices: $J_{\alpha \beta} = (J_{\alpha\alpha} + J_{\beta\beta})/2$. We note that
$J_{\alpha\alpha}$, with entries $(J_{\alpha\alpha})_{mn} = \partial F_m/\partial
\Lambda_n$, is the Jacobian of the system of equations $F_m(\bm \Lambda^\alpha)=0$
defined in Eq.~\eqref{eq:ebv}~\cite{claeys2017}. The normalization factor $A_\alpha$ is
also defined in terms of $J_{\alpha\alpha}$:
\begin{align}
  A_\alpha = \frac{1}{\sqrt{(-1)^{N_\alpha^{\rm eff} + M_\alpha}
  \det(J_{\alpha\alpha})}}, \label{eq:normalization}
\end{align}
where the positivity of the norm ensures that $A_\alpha \in \mathbb R$; we choose a gauge
in which $A_\alpha$ is strictly positive. A nonzero inner product naturally requires that
both states contain the same number of electrons.

The central result of this work is that an expression analogous to
Eq.~\eqref{eq:innerproduct} can be used to calculate matrix elements of the fermionic
creation and annihilation operators. For the matrix element of
$\cop^\dagger_{\kappa\sigma_\kappa}$ between states $\ket{\Psi_\alpha}$ and
$\ket{\Psi_\beta}$, we first define the quantity
\begin{align}
  \mathcal I^{\kappa \sigma_\kappa}_{\alpha\beta} = A_\alpha A_\beta \Sigma^{\kappa
  \sigma_\kappa}_{\alpha\beta} {(-1)^{N_\alpha^{\rm eff} + M_\alpha}\det
  J^\kappa_{\alpha\beta}}, \label{eq:Idef}
\end{align}
allowing the matrix element to be expressed explicitly:
\begin{align}
  {\bra{\Psi_\alpha} \cop_{\kappa \sigma_\kappa}^\dagger \ket{\Psi_\beta}} =
  \begin{cases}
    \mathcal I^{\kappa \sigma_\kappa}_{\alpha\beta} & \text{ for } \C_\alpha = \C_\beta
    \cup \{ (\kappa, \sigma_\kappa) \}, \\
    \sigma_\kappa \mathcal I^{\kappa \bar \sigma_\kappa}_{\beta\alpha} & \text{ for
    } \C_\beta = \C_\alpha \cup \{ (\kappa, \bar \sigma_\kappa) \}, \\
    0 & \text{ otherwise,}
  \end{cases} \label{eq:matrix_element_c}
\end{align}
where $\sigma_\kappa = \pm 1$ for spin up/down, $\bar\sigma_\kappa$ denotes the opposite spin, and
\begin{align}
  \Sigma^{\kappa \sigma_\kappa}_{\alpha\beta} = \bra\vac \left[\prod_{(n,\sigma_n) \in
  \C_\alpha} \cop_{n \sigma_n}^\dagger\right]^\dagger \cop_{\kappa \sigma_\kappa}^\dagger
  \!\!\!\! \prod_{(m,\sigma_m) \in \C_\beta} \!\!\! \cop^\dagger_{m \sigma_m} \ket\vac ,
\end{align}
with $\Sigma^{\kappa\sigma_\kappa}_{\alpha\beta} \in \{0, \pm 1\}$. In addition to the
conditions in Eq.~\eqref{eq:matrix_element_c}, the matrix element also vanishes unless
$n_{G, \alpha} = n_{G, \beta} + 1$.

Moreover, $J^{\kappa}_{\alpha\beta}\in \mathrm{Mat}(\min\{N_\alpha^{\rm eff},
N_\beta^{\rm eff}\}, \mathbb R)$ is defined for $m,n \in \nB $, with $\nB = \nB_\alpha
\cap \nB_\beta$:
\begin{align}
  (J_{\alpha \beta}^\kappa)_{mn} =
  \begin{cases}
    \ebv_m^\alpha + \ebv_m^\beta - 1 + \frac g2 \!\!\! \displaystyle \sum_{\ell \in \nB
    \setminus \{m\} } \! \! Z_{\ell m} & \text{ for } m = n; \\
    \frac g2 Z_{mn} & \text{ for } m \neq n.
  \end{cases}\label{eq:Jkappa}
\end{align}
This expression is derived in App.~\ref{app:matrix_element_c}. Matrix elements of the
annihilation operator follow by Hermitian conjugation, $\bra{\Psi_\alpha}
\cop_{\kappa\sigma_\kappa} \ket{\Psi_\beta} = \bra{\Psi_\beta}
\cop^\dagger_{\kappa\sigma_\kappa} \ket{\Psi_\alpha}^*$, and the empty determinant is
defined as unity, so ``arbitrary seniority'' includes the fully blocked cases. The sign
factor $(-1)^{N_\alpha^{\rm eff} + M_\alpha}$ in Eq.~\eqref{eq:Idef} depends only on
$\ket{\Psi_\alpha}$ (for $\C_\alpha = \C_\beta \cup \{(\kappa, \sigma_\kappa)\}$). This
follows from the matrix element being related to an inner product of the form given in
Eq.~\eqref{eq:innerproduct} in which only the Bethe-ansatz part of $\ket{\Psi_\alpha}$ is
on shell (i.e., the rapidities $u_{\nu}^\alpha$ satisfy the Richardson equations with
blocked levels $\B_\alpha$). The analogous statement holds for $\C_\beta = \C_\alpha \cup
\{(\kappa, \bar\sigma_\kappa)\}$, with $\alpha \leftrightarrow \beta$ and $\sigma_\kappa
\leftrightarrow \bar\sigma_\kappa$.

In the following sections, we use this identity to calculate matrix elements of tunneling
terms between subsystems in the product-state basis.

\subsection{Numerical calculation of the EBVs}
As in Ref.~\cite{elaraby2012}, the numerical solver is initialized at $g = 0$,
where $\ebv_m^\alpha\big\vert_{g = 0} = \langle{\hat S^z_m}\rangle_\alpha + 1/2$.
This is a convenient initialization point because the eigenstates of \hrg are $\hat
S_m^z$ eigenstates at $g=0$.
The solver consists of two steps: (1) a prediction step, in which we estimate
$\ebv_{m}^\alpha(g + \mathrm{d}g)$ given $\ebv_{m}^\alpha(g)$, and (2) a refinement step,
in which Eq.~\eqref{eq:ebv} is enforced using a multidimensional Newton method.
The step size $\mathrm{d}g$ is adjusted adaptively, which is especially important when
the Jacobian of Eq.~\eqref{eq:ebv} becomes nearly singular.

In our implementation, we use first-order prediction followed by Newton refinement
with adaptive step-size control.
For certain excited states the continuation passes through regions where the EBV
Jacobian is poorly conditioned at standard floating-point precision; this conditioning
limitation is resolved by using arbitrary-precision arithmetic implemented with
\texttt{python-flint}~\cite{flint, python_flint, johansson2017}.
Together, arbitrary-precision arithmetic and the adaptive-step continuation allow
reliable calculation of all RG eigenstates required in the present calculations,
each specified by a $g=0$ occupation configuration.

The EBV equations were solved using 200-digit precision with 40 additional guard digits
and a residual tolerance of $10^{-32}$. For $N = 30$ grain levels and $n_F = 30$ total
electrons at $k^{\rm eff} \leq 1$, the full EBV calculation required approximately 116 s,
whereas a single sparse eigensolve took approximately 6 ms. For a fixed grain spectrum,
pairing strength, and projected basis, the EBV data are independent of the tunneling and
charging parameters and can be saved permanently for reuse. Therefore, repeated sparse
diagonalizations and evaluation of the projected effective Hamiltonians dominate the cost
of extensive parameter sweeps, rather than EBV continuation or determinant evaluation.
For further details on the numerical implementation, see the Zenodo repository~\cite{zenodocode}.

\subsection{State selection \& projection} \label{sec:state-selection}

The exponential growth of the Hilbert space restricts practical calculations to the
low-energy sector. We implement this restriction by relating the $g=0$ level
occupations to the energy asymptotics at large $g$.

For this purpose, we employ the algorithm introduced in Ref.~\cite{roman2003} [see
App.~\ref{app:algo}], which assigns to each level-occupation configuration at $g = 0$ an
integer $k_\alpha$ that characterizes the excitation degree relative to the ground state.
For seniority-zero states with fixed particle number, $k_\alpha = 0$ for the ground state
and $k_\alpha = 1$ for the first band of excited states. Extending this definition, we define
\begin{align}
  k^{\rm eff}_\alpha \equiv k_\alpha + \vert \B_\alpha \vert/2, \quad 0 \leq
  2k^{\rm eff}_\alpha \leq \min(n_{G,\alpha}, 2N - n_{G,\alpha}),
\end{align}
as the effective excitation degree that accounts for the blocked levels and is
half-integer for odd $n_{G,\alpha}$. For even $n_{G,\alpha}$, $2k^{\rm eff}_\alpha = 2k_{\alpha} +
\vert \B_\alpha \vert$ measures the excitation degree relative to the ground state, counting both
pair excitations and the unpaired electrons occupying the blocked levels. The large-$g$
asymptotic behavior of the energy of a given state is determined by $k_\alpha^{\rm
eff}$~\cite{roman2003}:
\begin{align}
  E_\alpha(g) - E_\alpha(0) \sim -g\left[ s_\alpha(s_\alpha+1) - m_\alpha(m_\alpha+1)
  \right] + \mathcal O(1),
\end{align}
where $m_\alpha = (N - n_{G, \alpha})/2$ and $s_\alpha = (N - 2k^{\rm eff}_\alpha)/2$.
Here, $s_\alpha$ is the strong-coupling total-pseudospin label\footnote{For
  nondegenerate $\epsilon_m$, the kinetic term does not commute with $\hat{\bm S}^2_{\rm
  tot}$, so a finite-$g$ eigenstate is not an exact pseudospin eigenstate; $s_\alpha$
labels the multiplet reached by continuation.} of the state $\ket{\Psi_\alpha}$, and
$m_\alpha$ is minus the $z$ component of the total pseudospin.
This motivates a minimal labeling scheme for eigenstates of \hrg:
\begin{align}
  \ket{\Psi_\alpha} = \prod_{(b, \sigma_b) \in \C_\alpha} \cop^\dagger_{b\sigma_b}
  \ket{n_{G, \alpha}, k_\alpha, a_{\alpha}; \B_\alpha}, \label{eq:grain_wavefunction}
\end{align}
where $k_\alpha$ denotes the excitation degree with respect to the ground state,
$a_{\alpha}$ denotes the state index within the excited subspace, $\B_\alpha$ denotes the
blocked levels, and $\sigma_b$ denotes the spin of each blocked level.
The Bethe-ansatz part of the wave function $\ket{n_{G, \alpha}, k_\alpha,
a_{\alpha}; \B_\alpha}$ does not depend on the spins of the blocked levels and therefore
depends on $\B_\alpha$ rather than $\C_\alpha$.
The state index satisfies $0 \leq a_\alpha \leq \ydim(N^{\rm eff}_\alpha,
k_\alpha) - 1$, with~\cite{roman2003}
\begin{align}
  \ydim (N, k) = \binom Nk - \binom N{k-1}, \label{eq:ydim}
\end{align}
where $\binom N{-1}=0$ and $0 \leq k \leq \min(M, N-M)$ at fixed pair number $M$. At
fixed particle number, the number of states with excitation degree $k^{\rm eff}$ is
\begin{align}
  D(N, k^{\rm eff}) = \frac{N+1 - 2k^{\rm eff}}{N+1-k^{\rm eff}}\binom{2 N+1}{2 k^{\rm
  eff}}. \label{eq:exdim}
\end{align}
$\ydim(N, k)$ counts the number of seniority-zero states with excitation degree
$k$, whereas $D(N, k^{\rm eff})$ counts all states with excitation degree $k^{\rm eff}$
regardless of seniority.

We therefore choose to project the Hamiltonians in the following sections onto the
low-energy subspace ($k^{\rm eff} \leq 1$), reducing the scaling of the retained basis
from exponential to quadratic in the system size: at interior filling the retained basis
has dimension $2N(2N+1)$ (see App.~\ref{app:dim_ret}).
The projection is controlled when the single-particle gap $\Delta$ is large compared with
the coupling between the retained and discarded subspaces.
Denoting the tunneling amplitude by $\lambda$ (as introduced below), the
relevant control parameter is the coupling of the target low-energy states to
the discarded subspace divided by their energy separation; individual matrix
elements obey $|\langle \Psi_\alpha | \cop^{(\dagger)} |
\Psi_\beta\rangle| \leq 1$, but collective combinations of the $N$ tunneling
channels can be enhanced (a normalized collective orbital gives
$\sum_m \cop_{m\sigma} = \sqrt N\, \hat a_\sigma$), so $\lambda < \Delta$ is a
practical guideline rather than a sufficient bound.

Physically, the $k^{\rm eff} \leq 1$ truncation omits higher excited states from the
Hilbert space, whose effects on the low-energy physics are controlled by the spectral gap.

In the calculations presented here, EBVs were efficiently obtained for every state in
the retained Hilbert space. The growth of this basis ultimately limits the accessible
system sizes.
We use $N = 30$ single-particle levels around half filling.
The benchmarks below use $g = 1.2d$, i.e.\ $\Delta \approx 16.1d$ --- a strongly paired
mesoscopic regime; the effective-Hamiltonian method itself remains applicable, with the
same control parameters, toward the $\Delta \sim d$ regime emphasized in the introduction.

\subsection{Effective Hamiltonians}

The Hamiltonians appearing in the following sections are nonintegrable due to the
presence of single-electron tunneling. Nevertheless, we exploit the integrability of the
Richardson--Gaudin model by constructing effective Hamiltonians in a projected product
basis $\ket{\Psi} \otimes \ket{D}$, where $\ket{\Psi}$ denotes the grain component of the
wave function, defined as in Eq.~\eqref{eq:grain_wavefunction}, and $\ket{D}$ denotes the
dot part of the wave function. The dot basis consists of $\ket0$, $\ket\downarrow =
\d_\downarrow^\dagger \ket0$, $\ket\uparrow = \d_\uparrow^\dagger \ket0$, and $\ket2 =
\d_{\uparrow}^\dagger\d_{\downarrow}^\dagger \ket0$, denoting an empty dot, a spin-down
singly occupied dot, a spin-up singly occupied dot, and a doubly occupied dot,
respectively. We denote the total number of electrons in the state as $n_F = n_G + n_D$.
The fermionic operators satisfy the canonical anticommutation relations:
$\{\cop_{m\sigma},\cop_{n\tau}^{\dagger}\}=\delta_{mn}\delta_{\sigma\tau}$ and
$\{\d_{\sigma},\d_{\tau}^{\dagger}\}=\delta_{\sigma\tau}$; all other anticommutators
vanish, including those between grain and dot operators.

The matrix elements of the effective Hamiltonians are given by
\begin{align}
  H_{\alpha\beta} = (\bra {\Psi_\alpha} \otimes \bra{D_\alpha}) \hat H
  (\ket{\Psi_\beta}\otimes \ket{D_\beta}),
\end{align}
where the chosen fermionic ordering places the grain operators before the dot
operators; concretely, in the
product basis a dot operator acting past the grain factor carries the string
$(-1)^{\hat n_G}$.
As discussed above, the Hilbert space is projected onto the low-energy subspace with
$k^{\rm eff}_\alpha \leq 1$.

For the data presented in Figs.~\ref{fig:anderson}, \ref{fig:conductivity}, and
\ref{fig:flux}, the effective Hamiltonians are diagonalized using the sparse linear
algebra routines provided by the \texttt{SciPy} library~\cite{scipy}.

\subsection{Verification}
The method is tested by comparing the eigenvalues of the effective Hamiltonians with
those obtained by exact diagonalization of the full Hamiltonians for small system sizes,
for which the entire Hilbert space ($2k^{\rm eff} \leq \min(n_G, 2N-n_G)$) can be
included (see Ref.~\cite{zenodocode}). Under these conditions the results agree to
numerical precision. Additionally, the Bethe-ansatz wave function can be explicitly
constructed for sufficiently small systems [using Eq.~\eqref{eq:bethe_ebv}], allowing us
to compare the matrix-element expression of Eq.~\eqref{eq:matrix_element_c}, including
all sign factors, with the exact matrix elements. The resulting matrix elements agree to
machine precision for both cases of Eq.~\eqref{eq:matrix_element_c}, for all seniorities
and blocked-level configurations. Finally, when the low-energy projection $k^{\rm eff}
\leq 1$ is employed, the observed relative error in the benchmarked energies is of order
$(\lambda/\Delta)^2$ in the weak-tunneling limit; this is quantified in
App.~\ref{app:convergence}.


\vspace{2ex}
\section{Superconducting Anderson Model}\label{sec:anderson}

Using the techniques described in the previous section, we construct a projected
Richardson--Gaudin--Anderson (RGA)~\cite{pavesic2021, pavesic2022, baran2024, balatsky2006,
anderson1961} Hamiltonian from exact matrix elements between RG eigenstates:
\begin{align}
  \hrga = \hrg + \hhop + \hcharge,\label{eq:anderson}
\end{align}
where the tunneling Hamiltonian is defined as
\begin{align}
  \hhop = - \lambda \sum_{m = 1}^{N}\sum_{\sigma = \uparrow,\downarrow} \left(
  \cop^\dagger_{m\sigma} \d_\sigma + \mathrm{H.c.}\right).
\end{align}
The operator $\d_\sigma$ annihilates an electron on the dot with spin $\sigma$, and
$\lambda$ denotes the tunneling amplitude.
Finally, the charging Hamiltonian is defined as
\begin{align}
  \hcharge = U\left(\hat n_D - n_{0, D} \right)^2 + V\left(\hat n_G -
  n_{0, G} \right)^2, \label{eq:hcharge}
\end{align}
where $\hat n_D = \sum_{\sigma=\uparrow,\downarrow} \d_\sigma^\dagger \d_\sigma$ is the
dot number operator.
Physically, the charging terms arise from the capacitive coupling of the grain and dot to
gate electrodes; $U$ and $V$ set the charging energies, and $n_{0,D}$ and $n_{0,G}$
are the corresponding dimensionless gate-induced offsets
\footnote{
  Note that $U$ is a charging-energy coefficient rather than the conventional
  Anderson--Hubbard repulsion: $U(\hat n_D - n_{0,D})^2 = 2U \hat n_{D\uparrow}\hat
  n_{D\downarrow} + U(1-2n_{0,D})\hat n_D + \mathrm{const}$, i.e.\ $U_{\rm Hubbard} = 2U$
  when comparing scales to the Anderson-model literature.
}.

Superconducting Anderson impurity models have been studied extensively and are well known
to host subgap states in the single-particle excitation (SPE) spectrum~\cite{martinrodero2011}.
Throughout, SPE energies plotted above (below) zero are the addition (removal)
energies $\hbar \omega_+ = E_m(n_F+1) - E_0(n_F)$ ($\hbar \omega_- = -\,[E_m(n_F-1) -
E_0(n_F)]$), referenced to a reservoir chemical potential $\mu = 0$.
In the absence of charging, the dot experiences induced superconducting pairing through
the coupling to the grain, whereas for $U,V > 0$ this proximity effect competes with
local-moment formation.
Thus, by tuning $U$ and $V$, the system continuously transitions from a superconducting
grain coupled to a proximitized dot to a system that effectively behaves as a double quantum dot.

\subsection{Ground-state analysis for a singly occupied dot}
In this section, we set $n_{0, D}=1$ and consider the regime $U \gtrsim \Delta$.

\begin{figure*}
  \centering
  \includegraphics[width=\linewidth]{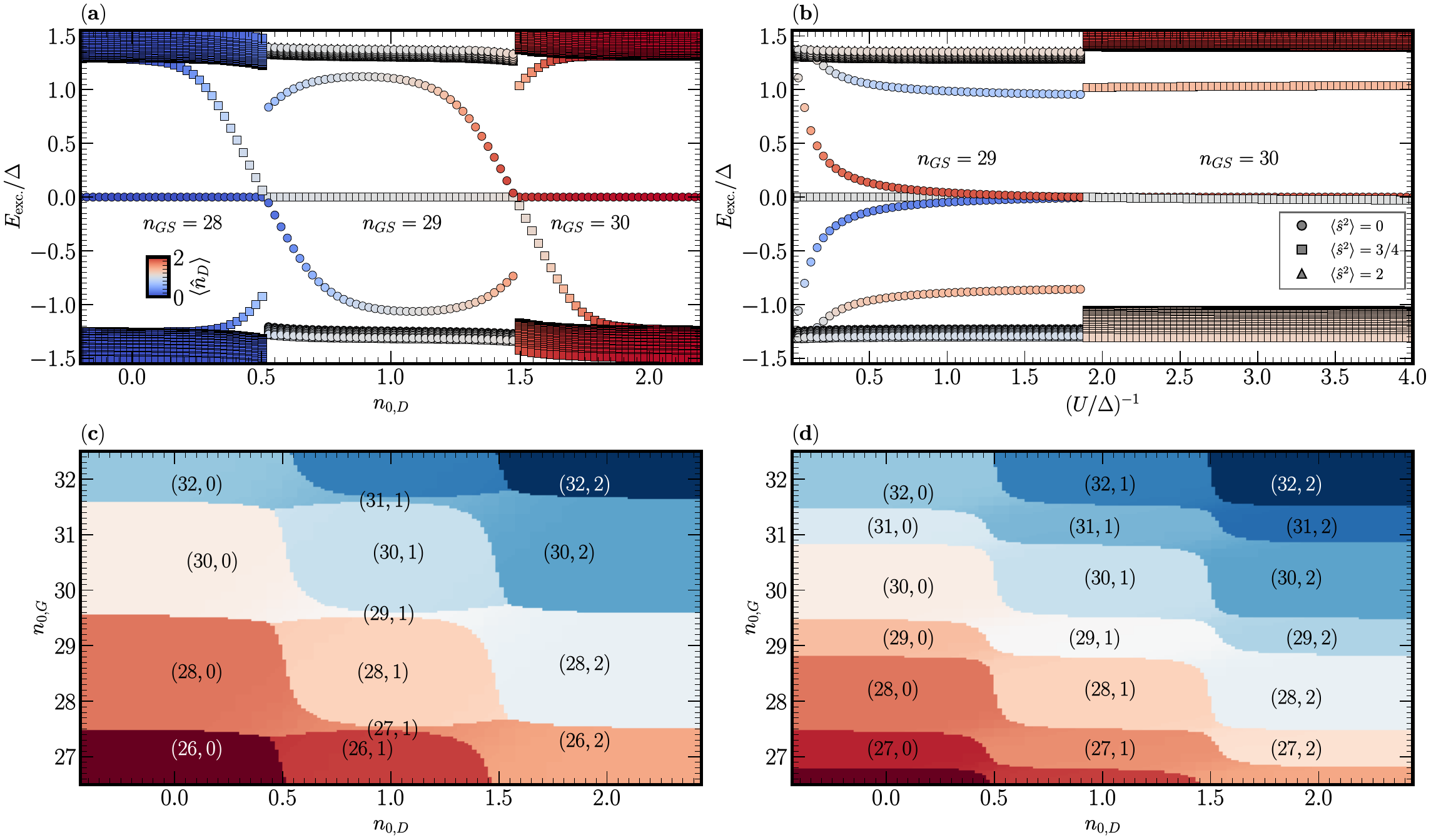}
  \caption{
    Panels (\andersonsubgap, \andersonphasetransition) show the single-particle
    excitation energies $E_{\rm exc.}$ of the superconducting Anderson model, with
    prominent subgap excitations.
    The excitation energies above zero in (\andersonsubgap,
    \andersonphasetransition) are obtained by adding a single electron to the ground
    state, whereas those below zero are obtained by removing a single electron.
    Varying the gate offset $n_{0, D}$ or interaction $U$ induces parity-changing
    transitions between doublet ($n_{F, GS}=29$) and spin-singlet ($n_{F, GS}=28,30$)
    ground states, where $n_{F, GS}$ denotes the number of electrons in the ground state.
    The marker color indicates dot occupation $\langle \hat n_D \rangle$, while the
    marker shape encodes total spin [see legend in (\andersonphasetransition)].
    The markers at zero energy display the dot occupation and total spin of the
    ground state.
    For even $n_{F, GS}$ near $n_{0, D} \approx 1$, the dot electron is screened by
    an unpaired grain electron, forming a spin singlet.
    Charge-stability diagrams emerge by varying $n_{0, D}$ and $n_{0, G}$:
    strong charging ($V$ large, \andersonstrong) resembles a double quantum dot, while
    weak charging ($V$ small, \andersonweak) shows superconducting behavior favoring even
    grain occupation.
    Tuples $(\langle \hat n_G \rangle, \langle \hat n_D \rangle)$ denote
    occupations, with the colors encoding $3\langle \hat n_G \rangle + \langle \hat
    n_D \rangle$, which uniquely labels the integer occupation sectors.
    Here, $N=30$, $n_F=26,27,\ldots,34$, $g=1.2d$, $\Delta=16.1d$. In
    (\andersonsubgap), $U=3\Delta$, $V = 0.26\Delta$, $\lambda=0.1\Delta$, and
    $n_{0, G}=29.5$; (\andersonphasetransition) uses the same parameters with
    $n_{0, D}=1$ and variable $U$.
    In (\andersonweak), $U=2.5\Delta$, $V=0.76\Delta$, and $\lambda=0.12\Delta$; in
    (\andersonstrong), $U=2.5\Delta$, $V=3\Delta$, and $\lambda=0.12\Delta$.
  }\label{fig:anderson}
\end{figure*}

Although the dot is approximately singly occupied, the even-parity grain-dot ground state
is a singlet.
This is because the spin of the electron on the dot is screened by an unpaired electron
in the superconducting grain:
\begin{widetext}
  \begin{align}
    \ket{\Psi_{\rm even}} \sim & \sum_{i=1}^N W_{1, i}^{\rm even} \left(
      \cop_{i,\uparrow}^\dagger \d_\downarrow^\dagger - \cop_{i,\downarrow}^\dagger
    \d^\dagger_{\uparrow}\right) \ket{n_{F}-1,0,0;\{i\}} \otimes\ket0
    + W_{0}^{\rm even} \ket{n_{F},0,0;\emptyset} \otimes \ket 0 + W_2^{\rm
    even} \ket{n_{F}-2,0,0;\emptyset}\otimes \ket2,\label{eq:wf_even}
  \end{align}
  where the basis $\ket{G} \otimes \ket{D}$ consists of product states of grain states
  $\ket G$ and dot states $\ket D$.
  The coefficients $W^{\rm even}_0$ and $W^{\rm even}_2$ describe the dominant
  amplitudes of configurations with an empty and doubly occupied dot, respectively;
  increasing $U$ suppresses charge fluctuations away from single occupancy and reduces
  the combined weight of these configurations.
  The terms in $\ket{\Psi_{\rm even}}$ proportional to $W^{\rm even}_{1,i}$ have
  singlet spin structure --- the unpaired electron in the grain screens the dot spin
  [see white circular markers in Fig.~\ref{fig:anderson}(\andersonsubgap)]; the
  terms proportional to $W^{\rm even}_0$ and $W^{\rm even}_2$ are also
  spin singlets due to the absence of unpaired electrons.
  Therefore, the competition between superconductivity and local-moment formation is
  encoded in the summed charge-sector weights: when $2\sum_i \vert W^{\rm
  even}_{1,i} \vert^2 \gg \vert W^{\rm even}_{0}\vert^2 + \vert W^{\rm even}_{2} \vert^2$
  (the factor $2$ counting the two spin components of the singlet), local-moment
  formation is dominant, and in the opposite limit the grain's preference for even
  occupation dominates.

  In contrast, in the odd-parity sector the ground states are of the form
  \begin{align}
    \ket{\Psi_{\rm odd, \sigma}} \sim W^{\rm odd}_{1} \ket{n_{F} - 1,0,0;\emptyset}
    \otimes \ket \sigma
    + \sum_{i=1}^N W^{\rm odd}_{0, i} \cop^\dagger_{i, \sigma}\ket{n_{F}, 0,0;
    \{i\}} \otimes \ket 0
    + \sum_{i=1}^N W^{\rm odd}_{2, i} \cop^\dagger_{i, \sigma} \ket{n_{F} - 2, 0,0;
    \{i\}} \otimes \ket 2,\label{eq:wf_odd}
  \end{align}
\end{widetext}
where the magnitudes of the coefficients $W^{\rm odd}_{0, i}$ and $W^{\rm odd}_{2,
i}$ decrease as $U$ increases [see Fig.~\ref{fig:anderson}(\andersonphasetransition)].
The dominant components of the numerically obtained wave functions in
Eqs.~\eqref{eq:wf_even} and~\eqref{eq:wf_odd} closely resemble the variational
ans\"{a}tze used in Ref.~\cite{lim2008}.

For weak $U$, the odd-parity ground-state doublet has an unpaired spin delocalized
between the dot and grain.
Increasing $U$ suppresses fluctuations away from $\mean{\hat n_D} = 1$ and thus
localizes the spin on the dot.

\subsection{Charge-stability diagram}
In the RGA model, the competition between superconducting pair correlations and Coulomb
charging gives rise to complex physics.
In particular, for weak charging on the grain [see
Fig.~\ref{fig:anderson}(\andersonweak)], the superconducting parity gap favors even
occupation of the grain.
On the other hand, if $V$ becomes comparable to or larger than $\Delta$, an odd electron
number in the grain can become energetically favorable near odd integer values of
$n_{0, G}$ [see Fig.~\ref{fig:anderson}(\andersonstrong)].

\subsection{Subgap states}
Depending on the parity of the ground state, the subgap states visible in
Fig.~\ref{fig:anderson} form either spin singlets or spin doublets.
In previous work (see Refs.~\cite{pavesic2021, pavesic2022}), these subgap states are
referred to as Yu-Shiba-Rusinov (YSR) states~\cite{balatsky2006, shiba1968, yu1965,
rusinov1969}; however, in this model, it is difficult to sharply distinguish between YSR
states and Andreev bound states~\cite{andreev1966}.
Therefore, we will refer to them as ``subgap states''.
As shown in Fig.~\ref{fig:anderson}(a,b), varying the gate offset $n_{0, D}$
and dot interaction strength $U$ can drive a subgap state through zero energy, marking a
ground-state crossing between opposite-parity sectors.
These transitions require electron exchange with a macroscopic reservoir on the
relevant timescales.
The spin on the dot is correlated with the spins in the superconducting grain, as can be
seen from the marker shapes and colors in Fig.~\ref{fig:anderson}(a,b).

\begin{figure*}[tb]
  \centering
  \includegraphics[width=\linewidth]{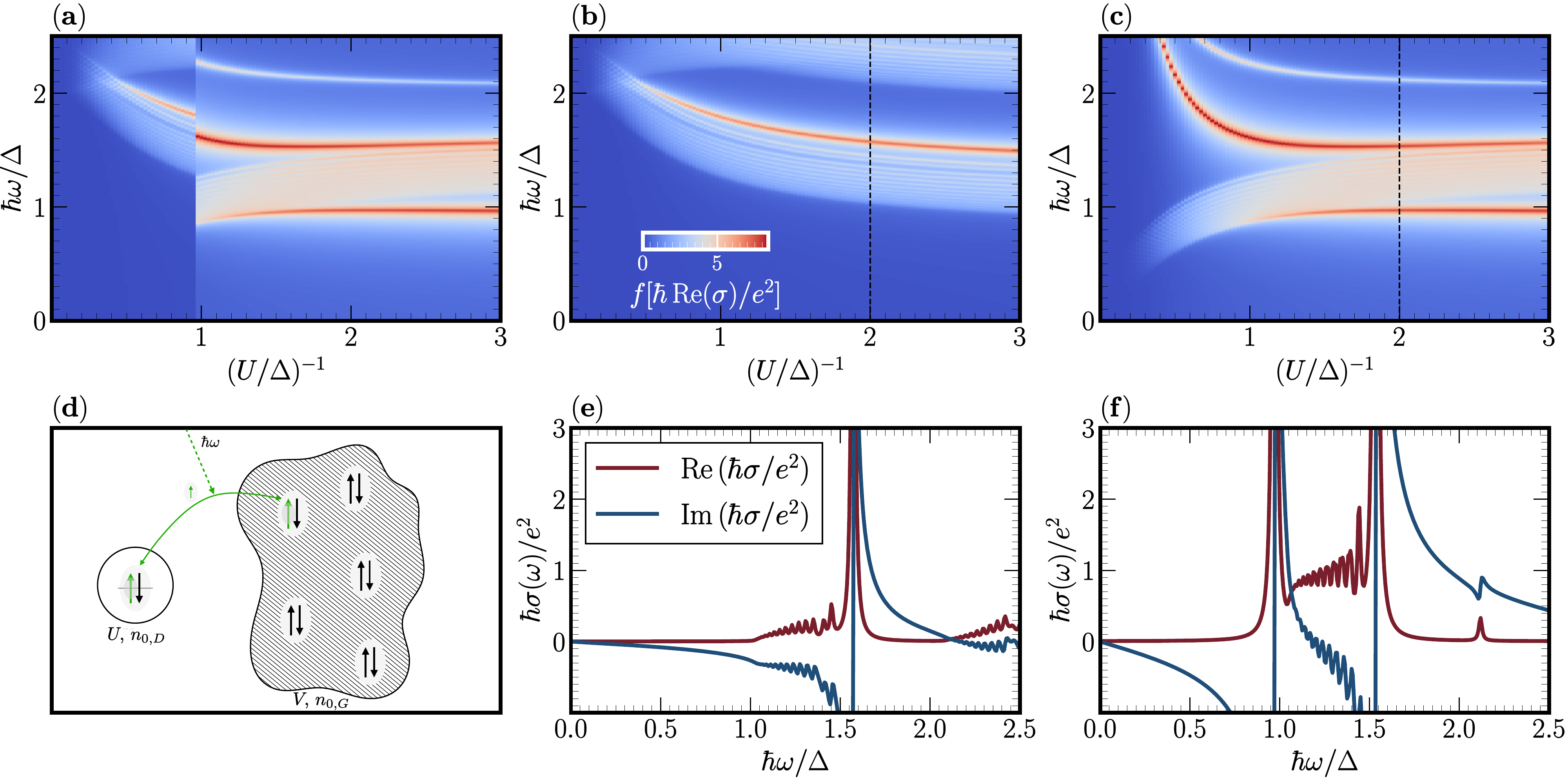}
  \caption{
    The optical conductivity exhibits peaks at the optically active transition
    frequencies of the system, where the strength of each peak is determined by the
    matrix element of the current operator. The transitions transfer electrons between
    the subsystems, as depicted in (\condfig). The optical conductivities in panels
    (\condodd) and (\condeven) are for a system with $n_{F, GS}=29$ and $n_{F, GS}=30$,
    respectively. Notably, in (\condodd), the dominant grain charge configuration has
    even parity, which, for the parameters shown, means that the lowest peak in the
    optical conductivity occurs around the single-particle gap. On the other hand, in
    panel (\condeven), in the small-$U$ limit, there is a pronounced subgap peak
    (associated with an excited state dominated by empty- and doubly-occupied-dot
    configurations). The vertical lines in panels (\condodd, \condeven) mark the value of
    $\Delta/U$ used for panels (\condgap, \condsubgap), respectively. When the phase
    transition from $n_{F, GS} = 29$ to $n_{F, GS} = 30$ is taken into account
    (\condboth), there is a sharp jump in the optical conductivity when a new set of
    transitions appears, including the transition to a subgap state. Panels (\condgap,
    \condsubgap) show the real and imaginary parts of the optical conductivity. The
    scaling function $f$ for panels (\condboth--\condeven) is defined as $f(x) = \log(1 +
    100 x)$, which makes smaller peaks more visible. Panel (\condfig) depicts a process
    in which a spin-up electron is transported between the subsystems in response to the
    applied drive. The spectral features are artificially broadened using $\hbar\omega
    \rightarrow \hbar\omega + i \eta$, with $\eta = \Delta / 100$. The parameters are $N
    = 30$, $g = 1.2d$, $\Delta = 16.1d$, $V = 0.4\Delta$, $n_{0, D} = 1$, $n_{0, G} =
    29.8$, $\lambda = 0.15\Delta$. See App.~\ref{app:sec:optical_conductivity} for further details.
  }\label{fig:conductivity}
\end{figure*}

\subsection{Optical conductivity}

Unlike single-particle spectra, the optical response probes excitations within a fixed
particle-number sector.
To study this response, we introduce a time-dependent modulation of the dot energy:
\begin{align}
  \hat{\tilde H}_{\rm cond.}(t) = \hrga + e \V(t)\hat n_D,
\end{align}
where the applied voltage $\V (t)$ is taken to have zero time average, and $e>0$ denotes
the elementary charge.
The modulation drives charge oscillations between the dot and the grain.
We transfer this modulation to the tunneling phase by moving to the rotating frame:
\begin{align}
  \hcond = \hat U^\dagger(t) \hat{\tilde H}_{\rm cond.}(t) \hat U(t) - i\hbar \hat
  U^\dagger(t) \partial_t{\hat U}(t),
\end{align}
where $\hat U(t) = \exp[-ie/\hbar \int_{-\infty}^t \nd t'\; \V(t')\, \hat n_D]$.
In the rotating frame, the Hamiltonian is
\begin{align}
  \hcond = \hrg + \hhopphase + \hcharge,
\end{align}
where the phase-dependent tunneling Hamiltonian is defined as
\begin{align}
  \hhopphase = - \sum_{m=1}^N\sum_{\sigma = \uparrow,\downarrow} \left( \lambda e^{i e
  A(t) / \hbar}\cop^\dagger_{m\sigma} \d_\sigma + \mathrm{H.c.}\right).
\end{align}
We define $A(t) = - \int_{-\infty}^t \nd t'\; \V(t')$, so that $eA(t)/\hbar$ is the
Peierls phase associated with the applied voltage.

Within linear-response theory, the resulting current $\mean{\curr(\omega)}$ between the
subsystems is proportional to the Fourier transform of the applied voltage:
\begin{align}
  \mean{\curr(\omega)} = \sigma(\omega) \V(\omega), \label{eq:ohm}
\end{align}
where $\curr(\omega) = -\mathcal F[ \partial_A\hhopphase ](\omega)$ and $\V(\omega)
= \mathcal F[ \V(t) ](\omega)$, and $\mathcal F[f](\omega)$ denotes the Fourier transform
of $f(t)$\footnote{We use $\mathcal F[f](\omega) \equiv \int \nd t \exp(i \omega t) f(t)$, and
  have the factor $(2\pi)^{-1}$ in the inverse transform. With this convention $\mathcal
  F[\partial_t f](\omega) = -i \omega \mathcal F[f](\omega)$; since $\partial_t A(t) = -
\mathcal V(t)$, this gives the factor $(i\omega)$ in Eq.~\eqref{eq:sigma}}. At zero
temperature, the optical conductivity is~\cite{defilippis2014, kubo1957}:
\begin{align}
  \sigma(\omega) = \frac{1}{i \omega} \left[ \Pi_0^{R}(\omega) + \mean{\hat K_0}_{0}
  \right], \label{eq:sigma}
\end{align}
where $\hat K_0$ is the diamagnetic term [$\hat K_0 \equiv -\partial_A^2 \hcond \vert_{A
= 0}$], and expectation values of the form $\langle \cdot \rangle_{0}$ are taken with
respect to the ground state.
\footnote{The quantity defined above is a lumped optical conductance (admittance)
with units of $e^2/\hbar$, not a bulk conductivity; no length or area conversion is implied.}

The retarded current-current correlation function $\Pi_0^{R}(\omega)$ is defined in the
Lehmann representation as~\cite{defilippis2014}:
\begin{align}
  \Pi_0^R(\omega) = -\sum_{m \neq 0} \vert \langle 0 \vert \curr_0 \vert m \rangle
  \vert^2 \frac{2\delta E_{m,0}}{(\hbar\omega + i 0^+)^2 - \delta E_{m,0}^2}, \label{eq:pi}
\end{align}
where the equilibrium current operator is defined as $\curr_0 \equiv -\partial_A
\hcond \vert_{A = 0} = - ie [\hrga, \hat n_D] / \hbar $, and the $m$-sum runs over
eigenstates of the equilibrium Hamiltonian.
The energy differences $\delta E_{m,0} = E_m - E_0$ are the excitation energies relative
to the ground state $\ket 0$.
For this finite, gapped system, the static paramagnetic and diamagnetic contributions
cancel, $\Pi_0^R(0) + \langle \hat K_0 \rangle_0 = 0$, removing the apparent $1/\omega$
singularity.
Consequently, $\sigma(\omega)$ vanishes as $\omega \rightarrow 0$, as expected for a
capacitive response\footnote{
  With the broadened numerical prescription $\hbar\omega \rightarrow \hbar\omega + i
  \eta$ used in Fig.~\ref{fig:conductivity}, the exact
  zero-frequency cancellation is recovered only as $\eta \rightarrow 0$; at finite $\eta$
  the broadened response has a small nonzero value at $\omega = 0$ of order $\mathcal O(\eta)$.
}.
At nonzero frequencies, the optical conductivity is suppressed below the first optically
active transition, but grows sharply near the resonance peaks.
At these resonances, the time-dependent dot energy drives coherent charge oscillations
between the subsystems, as depicted in Fig.~\ref{fig:conductivity}(\condfig).

By tuning $U$ [see Fig.~\ref{fig:conductivity}], we can not only shift the resonant
frequencies but also induce a discontinuous change in the optical conductivity.
This discontinuity occurs at a parity-changing ground-state transition at which a subgap
excitation becomes optically visible [see $U \approx \Delta$ in
Fig.~\ref{fig:conductivity}(\condboth)].
Such a parity-flipping transition implicitly assumes that the system can exchange a
single electron with an external reservoir, which supplies or absorbs the electron needed
to change the fermionic parity.
In a strictly isolated system with conserved particle number, this transition would not
occur, and the corresponding sharp discontinuity in the optical response would
therefore be absent, as shown in Fig.~\ref{fig:conductivity}(\condodd, \condeven)
for the odd- and even-parity sectors, respectively.

When $n_{0, D} = 1$ and the total electron number is odd, the dot is approximately
singly occupied in the ground state [see Fig.~\ref{fig:conductivity}(\condodd,\condgap)].
The excitations then predominantly involve either zero or two electrons on the dot,
together with an unpaired fermion in the grain [see the excited bands in
Fig.~\ref{fig:conductivity}(\condodd, \condgap)].
Increasing $U$ thus makes these excited states less favorable.
On the other hand, when the total electron number is even [see
Fig.~\ref{fig:conductivity}(\condeven, \condsubgap)], for small $U$, the ground state is a
superposition of configurations in which the dot is empty or doubly occupied.
The resulting ground state supports a spin-singlet subgap excitation.
As $U$ increases, both the empty and doubly occupied dot configurations become
energetically unfavorable, and the ground state acquires a single electron on the dot and
an unpaired electron in the grain.
The lowest neutral excitation energies approach zero as $U \rightarrow \infty$, while the optical
weights of the corresponding transitions vanish. In this limit the lowest \textit{optically active}
gap is finite.

\section{Flux loop}\label{sec:flux}

\begin{figure*}[!t]
  \centering
  \includegraphics[width=\linewidth]{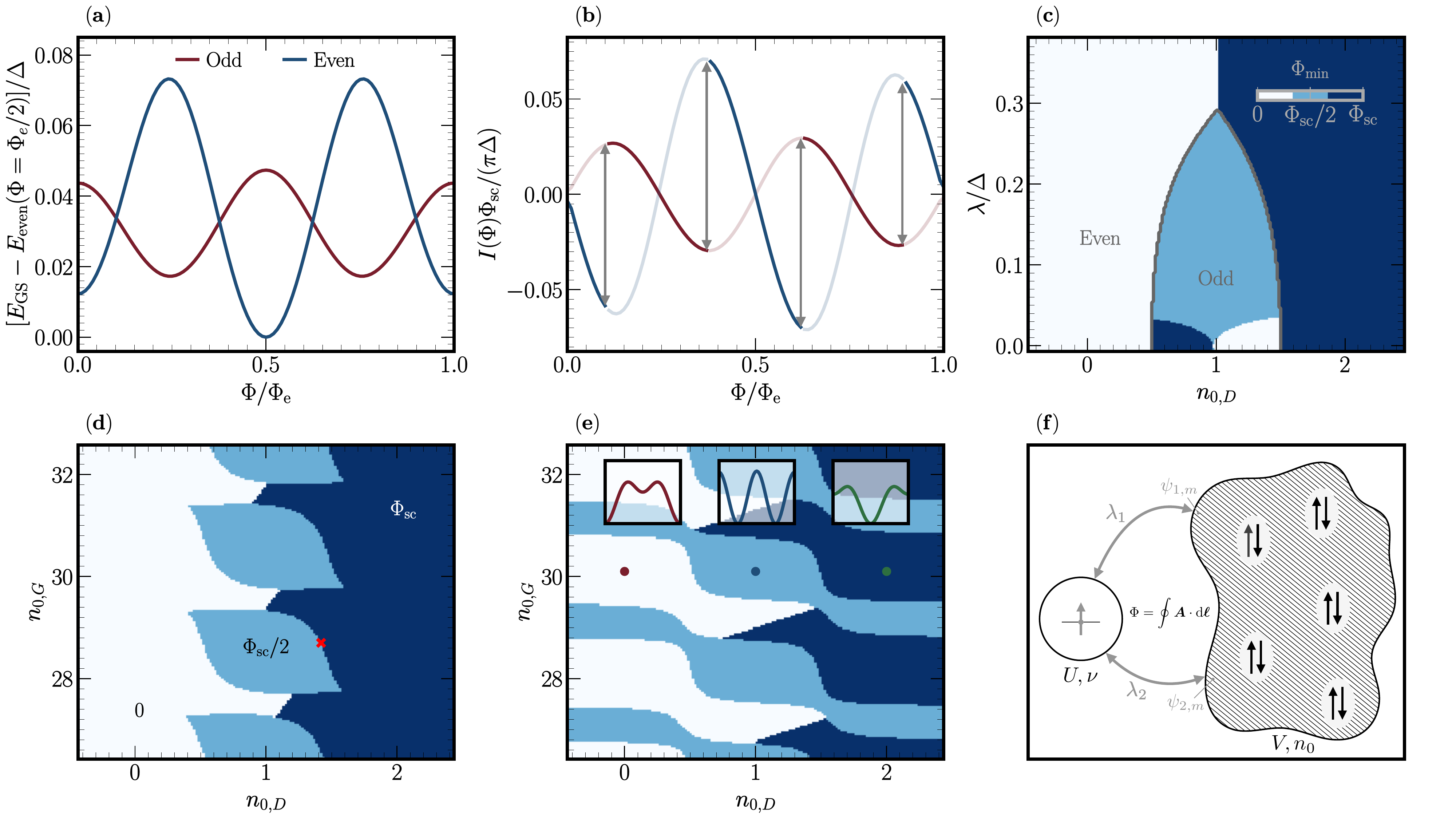}
  \caption{
    The energy-flux relation (\fluxepr) of the grain-dot system near a phase transition
    [red cross in (\fluxweak)] exhibits a characteristic even-odd structure. The spectrum
    is periodic in the electronic flux quantum, reflecting the presence of
    single-electron processes. In the odd-parity sector, the minimum lies at fractional
    multiples of $\Phi_{\rm sc} = h/(2e)$. However, if $\lambda \ll \Delta$ and the
    leading $\Phi_e = h/e$ harmonic is not suppressed, this harmonic becomes dominant.
    The energy minimum then approaches $\Phi = n\Phi_{\rm sc}$ with $n\in
    \mathbb Z$ [see (\fluxdome)]. When the even- and odd-parity energies cross, the
    ground-state parity can switch as a function of the flux $\Phi$, leading to
    discontinuities in the current-flux relation shown by the gray arrows in (\fluxcpr).
    The pale curves in (\fluxcpr) show the current–flux relations obtained by following
    the even- and odd-parity sectors separately, assuming that fermion parity is
    conserved. Moreover, tuning $n_{0, D}$ and $n_{0, G}$ shifts the relative energies of
    the parity sectors, thereby enabling $0$--$\pi$ transitions ($0$ and $\pi$ refer to
      the superconducting phase $\varphi = 2\pi\Phi/\Phi_{\rm sc}$ of the minimum,
    consistent with the Josephson $0$--$\pi$ transition of two-superconductor junctions).
    This is illustrated in (\fluxweak) and (\fluxstrong): the inline flux symbols mark
    the minimizing value of $\Phi$, namely $\Phi = 0$ or $\Phi = \Phi_{\rm sc}$ for an
    even total electron number and $ \approx \Phi_{\rm sc}/2$ for an odd one [compare
      with the charge-stability diagram in Fig.~\ref{fig:anderson}(\andersonweak,
    \andersonstrong)]. $N = 30$, $g = 1.2d$, $\Delta = 16.1d$, $\lambda = 0.12\Delta$,
    $U=2.5\Delta$. In (\fluxepr, \fluxcpr) $n_{0, D} = 1.42$, $n_{0, G} = 28.5$; in
    (\fluxdome) $n_{0, G} = 28.5$; in (\fluxepr--\fluxweak), $V=0.75\Delta$; in
    (\fluxstrong), $V=3\Delta$.
  }\label{fig:flux}
\end{figure*}

The model in Eq.~\eqref{eq:anderson} can be extended to describe a dot coupled to an
ultrasmall superconducting grain at two spatially separated contact points.
We model the two contacts by allowing the tunneling amplitudes to depend on the
grain-level index, thereby accounting for the fact that different grain wave functions
can have different relative signs at the two contact points.
Additionally, we choose a gauge in which the flux-dependent Peierls phase appears on one
of the two tunneling paths, as depicted in Fig.~\ref{fig:flux}(\fluxdepiction).
Thus, the Hamiltonian is defined as
\begin{align}
  \hrgaphi &= \hrg + \hhopflux + \hcharge.
\end{align}
The flux-dependent tunneling Hamiltonian is
\begin{align}
  \hhopflux =
  - \sum_{m=1}^N \sum_{\sigma = \uparrow,\downarrow} \left(\lambda_m(\Phi)
  \cop^\dagger_{m\sigma} \d_\sigma + \mathrm{H.c.}\right),
\end{align}
where $\lambda_m(\Phi) = \lambda_{1,m} + \lambda_{2,m}(\Phi)$:
\begin{align}
  \lambda_{1,m} = \lambda \psi_{1,m}, \qquad \lambda_{2,m}(\Phi) = \lambda \psi_{2,m}
  \exp[2 \pi i \Phi/\Phi_e]. \label{eq:fluxtun}
\end{align}
The overall tunneling scale is set by $\lambda$, and $\psi_{j,m} = \pm 1$ for $j=1,2$
encode the signs of the $m$th grain wave function at the two contact points [see
Fig.~\ref{fig:flux}(\fluxdepiction)].
$\Phi = \oint_{\Gamma} \bm A \cdot \nd \bm\ell$ is the flux through the loop, where
$\Gamma$ is the physical loop contour.
The single-electron flux quantum is defined as $\Phi_e = h/e$.
For simplicity, we set $\psi_{1,m} = 1$ and $\psi_{2,m} = (-1)^m$, which is a minimal
treatment of the two independent channels.

The two spatially separated tunneling paths give rise to Aharonov-Bohm
interference~\cite{aharonov1959}, making the spectrum sensitive to the enclosed flux.
Since the phase accumulated around the loop is set by the charge of the mobile
excitations, the energies are generally periodic in $\Phi_e$.
However, the relative strength of the $\Phi_e$ and $\Phi_{\rm sc} = h/(2e)$ harmonics is
parameter dependent.

For instance, in the $\lambda \ll \Delta$ limit, single-electron motion is generically
dominant. Provided the leading $\Phi_e$ harmonic is not suppressed, the energy minimum
lies near fluxes satisfying $\Phi \approx n \Phi_{\rm sc}$ for $n\in\mathbb{Z}$ [see
Fig.~\ref{fig:flux}(\fluxdome)].

With increasing $\lambda/\Delta$, the $\Phi_{\rm sc}$-periodic harmonic becomes more
prominent. Pair-transfer processes are expected to contribute to this harmonic in a
superconducting grain, but higher-order single-electron interference can also contribute,
particularly in the presence of charging. The harmonic alone does not distinguish these
contributions. Numerically, we find that the odd-parity minimum shifts to $\Phi \approx
\Phi_{\rm sc}/2$, whereas the even-parity minimum remains at either $\Phi=0$ or $\Phi =
\Phi_{\rm sc}$ [see Fig.~\ref{fig:flux}(\fluxdome)]. For the parameters of
Fig.~\ref{fig:flux}(\fluxdome), further increasing $\lambda/\Delta$ favors the
even-parity sector, as shown by the gray contour, which denotes the parity-flipping phase
boundary. The ``dome'' shown in Fig.~\ref{fig:flux}(\fluxdome) is a feature that also
appears in superconducting quantum-dot hybrids, as in Refs.~\cite{meng2009, lee2014,
maurand2012}. The value of $n_{0, G}$ used for Fig.~\ref{fig:flux}(\fluxdome) is close to
a value for which the system is particle-hole symmetric --- at this point, moving an
electron from the dot to the grain costs as much energy as moving an electron from the
grain to the dot. This accounts for the approximate symmetry of the parity-flipping
boundary under the transformation $n_{0,D} \rightarrow 2-n_{0,D}$. At effective
particle-hole symmetry, destructive interference suppresses the leading $\Phi_e$
harmonic at second order in $\lambda$.

When the remaining $\Phi_{\rm sc}$-periodic component is dominant, as in our numerical
results, this yields $E(\Phi) \simeq E(\Phi + \Phi_{\rm sc})$ up to corrections from
residual odd harmonics.

The second-order perturbative argument in App.~\ref{app:sec:pert} explains the
suppression of the $\Phi_e$ harmonic and the sign change of its coefficient between the
$n_{0,D}=0$ and $n_{0,D}=2$ regimes, accounting for the associated change in the location
of the energy minimum. The dominance of the $\Phi_{\rm sc}$ component and the odd-parity
minimum near $\Phi_{\rm sc}/2$ over the plotted range of $n_{0,D}$ are inferred from the
numerical energy-flux relations.

Finally, the persistent current through the dot can be calculated from
\begin{align}
  I(\Phi) = - \frac{\partial E_{\rm GS}}{\partial \Phi}, \label{eq:cpr}
\end{align}
and is similarly tunable by varying the system parameters.
Here, $E_{\rm GS}$ denotes the ground-state energy of the full system.
Specifically, the form of the current-flux relation depends on the parity of the ground
state and, near the parity-flipping phase transitions in
Fig.~\ref{fig:flux}(\fluxweak,\fluxstrong), can display discontinuities, as shown in
Fig.~\ref{fig:flux}(\fluxcpr)~\cite{martinrodero2011}.
Here, as in Sec.~\ref{sec:anderson}, these parity-flipping transitions occur only if the
system is allowed to exchange particles with a macroscopic reservoir.

\section{Conclusion}

We have developed a projected effective-Hamiltonian approach for hybrid mesoscopic
superconducting systems, in which the superconducting grain is treated exactly through
the Bethe-ansatz solution of the Richardson--Gaudin Hamiltonian.
Our central technical result is a compact determinant expression for matrix elements of
fermionic creation and annihilation operators, formulated entirely in terms of
eigenvalue-based variables.
Unlike previously available eigenvalue-based expressions for overlaps and selected form
factors, this result applies to single-electron processes between states of arbitrary
seniority, and therefore provides the missing ingredient needed to couple an integrable
superconducting grain to degrees of freedom that break integrability.

Using these matrix elements, we construct low-energy effective Hamiltonians in a basis of
Richardson--Gaudin eigenstates and impurity states.
The resulting method naturally works in the canonical ensemble, where particle-number and
parity effects are retained, while the low-energy projection reduces the otherwise
exponential Hilbert-space growth to a tractable problem.

The approximation is controlled by the retained-to-discarded coupling relative to the
corresponding energy separation, and its accuracy can be systematically improved by
enlarging the retained excitation subspace.

As a proof of principle, we applied the method to the superconducting Anderson model
consisting of an ultrasmall superconducting grain coupled to a single interacting level.
The approach captures the competition between pairing correlations, charging effects, and
local-moment formation, including even-odd effects, singlet-doublet transitions, and
subgap states with YSR-like character.
In addition, the same effective-Hamiltonian framework gives access to observables beyond
the equilibrium spectrum: we calculated the optical conductivity, which distinguishes
fixed-particle-number excitations from single-particle addition and removal spectra, and
studied flux-tunable grain-dot devices displaying interference effects, parity-dependent
current-flux relations, and $0$--$\pi$ transitions.

To our knowledge, the optical conductivity and flux-tunable interference spectra
studied here have not previously been calculated for a grain-dot system within the
canonical ensemble.

More broadly, the method provides a route for exploiting integrability in experimentally
motivated systems for which integrability is broken only by couplings to a small number
of additional degrees of freedom.
It can be extended directly to multiple quantum dots, several superconducting grains, and
more general tunneling geometries.

The matrix elements between RG eigenstates are also the operator content required to
embed a grain as a single exact site in a tensor-network calculation. At fixed
excitation-degree cutoff, the projection procedure yields a retained basis whose size
grows polynomially with the number of grain levels. Together with the determinant
expressions, this allows polynomial-cost construction of the site operators without
forming exponentially large configuration-space wave functions. The site operators follow
from the EBVs alone, obtained numerically from Eq.~\eqref{eq:ebv}, with normalization and
gauge fixed by Eq.~\eqref{eq:normalization} and a single coefficient of
Eq.~\eqref{eq:bethe_ebv}. Only the set of retained grain eigenstates must be truncated,
in direct analogy to the $k^{\rm eff}$ projection of Sec.~\ref{sec:state-selection}.
Since the grain then occupies a single site of a matrix-product chain, the attached
conduction degrees of freedom can be treated using DMRG --- in particular a long wire
coupled to the grain at one end --- with the grain allowed to fluctuate between the
retained charge sectors while the full system remains at fixed particle number. This
provides a number-conserving description of proximity-induced pairing.

It would also be valuable to generalize the determinant expressions to other
Richardson--Gaudin families and to higher pseudospin representations, which could enable
applications to systems with orbital or spatial degeneracies.
These extensions may make it possible to treat networks of mesoscopic superconducting
islands while retaining a microscopic, number-conserving description of pairing correlations.

\section{Code and data availability}

The data and code supporting the results reported in this article are publicly available
via Zenodo~\cite{zenodocode}.

\begin{acknowledgments}
  The authors would like to thank Maximilian H\"{u}nenberger and Sebastian Miles for
  useful discussions. This work was supported by the Swiss National Science Foundation
  and NCCR SPIN (Grant No.\ 51NF40-225153). D.L. acknowledges the Deanship of Research
  and the Quantum Center at KFUPM for the support received under Grant Nos.~CUP25102
  and~INQC2600, respectively. The authors acknowledge the use of OpenAI’s ChatGPT and
  Anthropic's Claude as auxiliary tools for technical discussions, coding assistance,
  verification of calculations, and manuscript preparation. All scientific results and
  conclusions were independently verified by the authors.
\end{acknowledgments}


\appendix

\begin{widetext}

  \section{Table of symbols}

  \bgroup
  \renewcommand{\arraystretch}{1.5}
  \begin{longtable}{lll}
    \toprule
    \textbf{Symbol} & \textbf{Definition} & \textbf{Description} \\
    \midrule
    $N$ & $\sim$ & number of single-particle levels (each of which is twofold degenerate
    due to spin).\\
    $\C_\alpha$ & $\sim$ & configuration set of blocked-level index-spin tuples in
    $\ket{\Psi_\alpha}$.\\
    $\B_\alpha$ & $\{b: (b, \cdot) \in \C_\alpha\}$ & set of blocked-level indices in
    $\ket{\Psi_\alpha}$.\\
    $\nB_\alpha$ & $\overline{\B_\alpha}$ & the set of unblocked levels in
    $\ket{\Psi_\alpha}$ (the complement of $\B_\alpha$).\\
    $\vert\B_\alpha\vert$ & $\sim$ & seniority of $\ket{\Psi_\alpha}$ (number of
    blocked levels).\\
    $N^{\rm eff}_\alpha$ & $N - \vert \B_\alpha \vert$ & number of unblocked levels in
    $\ket{\Psi_\alpha}$.\\
    $M_\alpha$ & $\sim$ & number of pairs in $\ket{\Psi_\alpha}$.\\
    $n_{G, \alpha}$ & $2M_\alpha + \vert \B_\alpha \vert$ & number of electrons in
    $\ket{\Psi_\alpha}$. \\
    $n_{F}$ & $n_G + n_D$ & number of electrons in the full grain-dot system.\\
    $k_\alpha$ & $N_g$ in Ref.~\cite{roman2003}& excitation degree with respect to the
    ground state (number of \textit{gaudinos}). \\
    $k_\alpha^{\rm eff}$ & $k_\alpha + \vert \B_\alpha \vert/2$ & effective excitation
    degree, including blocked levels. \\
    $s_\alpha$ & $N/2 - k_\alpha^{\rm eff}$& strong-coupling total-pseudospin
    label of $\ket{\Psi_\alpha}$. \\
    $m_\alpha$ & $N/2 - n_{G, \alpha}/2$ & minus the $z$ component of pseudospin of
    $\ket{\Psi_\alpha}$. \\
    $\epsilon_m$ & $md$ & single-particle energy of grain level $m$; the corresponding
    noninteracting pair energy is $2\epsilon_m$.\\
    $d$ & $\sim$ & single-particle level spacing.\\
    $\Omega$ & $Nd$ & bandwidth of the single-particle system.\\
    $\Delta$ & $\Omega / [2 \sinh(d/g)]$ & excitation gap in the thermodynamic limit.\\
    \hrg & Eq.~\eqref{eq:hrg} & reduced BCS Hamiltonian. \\
    $\ket{\Psi_{\alpha}}$ & Eqs.~\eqref{eq:bethe_ansatz} and~\eqref{eq:bethe_ebv} &
    Bethe-ansatz solution to \hrg.\\
    $u_\mu^\alpha$ & Eq.~\eqref{eq:richardson} & rapidity (half of the conventional
    Richardson pair energy).\\
    $\qm^\alpha$ & Eq.~\eqref{eq:conserved_charges} & conserved charge of \hrg. \\
    $\ebv_m^\alpha$ & Eq.~\eqref{eq:ebv} & eigenvalue-based variable. \\
    $\d^{(\dagger)}_{\sigma}$ & $\sim$ & fermionic annihilation (creation) operators on the dot.\\
    $\cop^{(\dagger)}_{m\sigma_m}$ & $\sim$ & fermionic annihilation (creation) operators
    on the grain.\\
    $\hat S_m^{+}, \hat S_m^-, \hat S_m^z$ & $\sim$ & Anderson pseudospin operators.\\
    $\Sigma_{\alpha\beta},\Sigma^{\kappa \sigma_\kappa}_{\alpha \beta}$ & $\sim$ & sign
    factors for inner products and matrix elements, respectively.\\
    $\V(t), \V(\omega)$ & $\sim$ & applied dot voltage in the time and frequency domains,
    respectively. \\
    $eA(t)/\hbar$ & $\sim$ & Peierls phase.\\
    $\sigma(\omega)$ & Eq.~\eqref{eq:sigma} & optical conductivity. \\
    $\curr(\omega)$ & Eq.~\eqref{eq:ohm} & current operator. \\
    $\curr_0$ & $\sim$ & equilibrium current operator. \\
    $\hat K_0$ & $-\partial_A^2 \hcond \vert_{A = 0}$ & diamagnetic operator. \\
    $\Pi_0^R$ & Eq.~\eqref{eq:pi} & retarded current-current correlation function. \\
    $J_{\alpha\beta}$ & Eq.~\eqref{eq:J} & matrix used to calculate inner products.
    $J_{\alpha\alpha}$ is the Jacobian of Eq.~\eqref{eq:ebv}. \\
    $J_{\alpha, P}'$ & Eq.~\eqref{eq:Jprime} & coefficient matrix of the EBV
    wave-function expansion (not a submatrix of $J_{\alpha\beta}$). \\
    $J_{\alpha\beta}^\kappa$ & Eq.~\eqref{eq:Jkappa}& modified $J$-matrix, used to
    calculate matrix elements of the creation/annihilation operators. \\
    $\Phi$ & $\oint \nd \bm \ell \cdot \bm A$ & applied flux. \\
    $\Phi_e$ & $h/e$ & single-electron flux quantum. \\
    $\Phi_{\rm sc}$ & $h/(2e)$ & superconducting flux quantum. \\
    $\lambda_m(\Phi)$ & Eq.~\eqref{eq:fluxtun} & flux-dependent tunneling amplitude. \\
    $I(\Phi)$ & Eq.~\eqref{eq:cpr} & persistent current. \\
    $g$ & Eq.~\eqref{eq:hrg} & pairing interaction strength ($g>0$
    attractive). \\
    $Z_{mn}$ & $(\epsilon_m - \epsilon_n)^{-1}$ & inverse
    level-splitting kernel. \\
    $A_\alpha$ & Eq.~\eqref{eq:normalization} & normalization of the
    (scaled) Bethe-ansatz state, chosen real and positive. \\
    $U, V$ & Eq.~\eqref{eq:hcharge} & dot and grain charging-energy
    coefficients ($U_{\rm Hubbard} = 2U$). \\
    $n_{0,D}, n_{0,G}$ & Eq.~\eqref{eq:hcharge} & dimensionless
    gate-induced charge offsets of dot and grain. \\
    $\hat n_D, \hat n_G$ & $\sim$ & dot and grain number operators. \\
    $\ydim(N,k)$ & Eq.~\eqref{eq:ydim} & seniority-zero asymptotic strong-coupling state
    multiplicity. \\
    $D(N,k^{\rm eff})$ & Eq.~\eqref{eq:exdim} & seniority-independent asymptotic
    strong-coupling state multiplicity. \\
    $\Delta_{\rm ML}$ & Ref.~\cite{matveev1997} & Matveev--Larkin parameter of
    an isolated grain, from the exact spectrum. \\
    \bottomrule
    \caption{Symbols, definitions, and descriptions of variables used in this work.}
  \end{longtable}
  \egroup

  \section{Conserved charges and the EBVs}
  \subsection{Energy}
  The Richardson--Gaudin Hamiltonian can be written in terms of the conserved charges:
  \begin{align}
    \hrg = \sum_{m} 2 \epsilon_m \hat Q_m + g \sum_{m \in \nB}\sum_{\ell \in \nB
    \setminus \{m\}} \left(\hat S^z_m \hat S^z_\ell - \frac14\right) - g \sum_{\ell}\hat
    S_{\ell}^{+} \hat S_{\ell}^{-}, \label{eq:app:hrg_cons}
  \end{align}
  where the double sum is restricted to the set of unblocked levels of a given state, $\nB$.

  Applying \hrg to an eigenstate, $\hrg \ket{\Psi_\alpha} = E_\alpha \ket{\Psi_\alpha}$,
  allows us to express the energies in terms of the EBVs:
  \begin{align}
    E_{\alpha} = \sum_{m} 2 \epsilon_{m} \ebv_{m}^\alpha - g M_\alpha (N^{\rm eff}_\alpha
    - M_\alpha + 1),
  \end{align}
  where
  \begin{align}
    \begin{split}
      & g\sum_{m \in \nB} \sum_{\ell \in \nB \setminus \{ m \}}
      \bra{\Psi_\alpha}\left(\hat S^{z}_{m} \hat S^{z}_{\ell} - \frac14\right)
      \ket{{\Psi_\alpha}}  \\
      & = g \sum_{m,\ell \in \nB} \bra{\Psi_\alpha}\left(\hat S^{z}_{m} \hat S^{z}_{\ell}
      - \frac14\right)\ket{\Psi_\alpha}= \frac g4 \left[(2M_\alpha-N_\alpha^{\rm eff})^2
      - (N_\alpha^{\rm eff})^2  \right] = -g M_\alpha (N_\alpha^{\rm eff}-M_\alpha) ,
    \end{split}
  \end{align}
  together with
  \begin{align}
    - g \sum_{\ell \in \nB} \bra{\Psi_\alpha} \hat S_{\ell}^{+} \hat S_{\ell}^{-}
    \ket{\Psi_\alpha} = -gM_\alpha,
  \end{align}
  resulting in the $-gM_\alpha(N_\alpha^{\rm eff}-M_\alpha+1)$ term.
  Here, the blocked-level energies are taken into account by using $\ebv_b^\alpha = 1/2$
  for $b \in \B_\alpha$.

  \section{Numerics and Projection}
  \subsection{State selection}\label{app:algo}
  We first describe the algorithm introduced in Ref.~\cite{roman2003} --- the excitation
  degree $k$ in this work corresponds exactly to their $N_g$.
  Each many-body $\hat S^{z}$-eigenstate (and hence eigenstate of \hrg at $g=0$) defines
  a unique path on a square lattice, constructed by scanning the noninteracting
  single-particle level-occupation pattern from lowest to highest: we step upward if a
  level is occupied by a pair and to the right otherwise (we ignore blocked levels for now).
  Restricting to states with $M$ pairs and $N\!-\!M$ unoccupied levels, each path starts
  at the origin and ends at the point $(N\!-\!M,M)$; all states share the same endpoints.

  The EBVs of every state (at $g>0$) are continuously connected to the energy-level
  occupations of an eigenstate at $g=0$.
  This correspondence permits the selection of relevant states to include in the calculations.

  The Young diagram associated with a state is obtained by comparing its path with that
  of the ground state.
  The two paths enclose a finite region wherever they differ; this region defines the
  Young diagram.
  Within this diagram, $k$ is given by the length of the longest southwest–northeast path
  (counting the boxes that are crossed along the path) that can be taken on the diagram.
  The ground state corresponds to $k=0$.
  For $k=1$, one obtains
  \begin{align}
    Y_1=\{\ldots,\ydiagram{1,1,1}, \ydiagram{1,1},\ydiagram{1},\ydiagram{2},
    \ydiagram{3}, \ldots\},
  \end{align}
  and for $k=2$,
  \begin{align}
    Y_2=\{\ydiagram{2,1},\ydiagram{3,1},\ydiagram{2,2},\ydiagram{2,1,1},\ldots\},
  \end{align}
  where $Y_{q}$ denotes the set of diagrams with $q = k$.

  Reference~\cite{roman2003} additionally gives the number $\ydim(N, q)$ of Young
  diagrams with $q=k$:
  \begin{align}
    \ydim (N, q) = \binom Nq - \binom N{q-1}. \label{eq:app:dNq}
  \end{align}

  When blocked levels are included, it is convenient to describe the excitation degree in terms of
  \begin{align}
    k^{\rm eff} = k + \vert \B \vert /2.
  \end{align}
  As discussed in the main text, $k^{\rm eff} \leq 1$ is our chosen production cutoff at
  the largest system sizes; enlarging it is systematically possible at increased cost.
  Thus, we explicitly give details on $k^{\rm eff} \leq 1$ below.

  \subsection{Even-parity sector}
  The ground-state occupation at $g=0$ is simply given by the Fermi sea:
  \begin{align}
    {\mathcal H}_{k^{\rm eff}=0, g=0}^{\rm even} = \{ \vert {\underbrace{\pair \pair
      \ldots \pair}_{M} \fs \underbrace{\unoccupied \unoccupied\ldots \unoccupied}_{N-M}}
    \rangle \},
  \end{align}
  where $\fs$ denotes the Fermi surface,  $\pair$ denotes a filled level, and
  $\unoccupied$ denotes an empty level.
  The levels are arranged from left to right in order of increasing energy.

  The seniority-zero states with $k^{\rm eff} = 1$ (of which there are $N-1$) are given
  by specific pair excitations of the ground state:
  \begin{itemize}
    \item Any one of the lowest $M-1$ pairs can be excited to the first level above the
      Fermi surface.
    \item The highest-energy pair can be excited to any level above the Fermi surface.
  \end{itemize}
  Additionally, there are seniority-two states with $k^{\rm eff}=1$ [of which there are
  $N(N-1)/2$, excluding spin degeneracy] which are obtained by blocking two distinct
  energy levels, and filling the lowest $M-1$ unblocked levels with pairs.

  Explicitly, for a small system ($N=6$ and $M=3$), the ground state is $\ket{GS} =
  \ket{\pair\pair\pair \fs \unoccupied\unoccupied\unoccupied}$.
  The set of first-excited states is given by
  \begin{align}
    \begin{split}
      {\mathcal H}_{k^{\rm eff} = 1, g = 0}^{\rm even} = \{
        \ket{{\unoccupied\pair\pair}\fs{\pair\unoccupied\unoccupied}},
        \ket{\pair\unoccupied\pair\fs\pair\unoccupied\unoccupied},
        \ket{\pair\pair\unoccupied\fs\pair\unoccupied\unoccupied},
        \ket{\pair\pair\unoccupied\fs\unoccupied\pair\unoccupied},
        \ket{\pair\pair\unoccupied\fs\unoccupied\unoccupied\pair},\\
        \ket{\pair\pair\unoccupied\fs\unoccupied\blocked\blocked},
        \ket{\pair\pair\unoccupied\fs\blocked\unoccupied\blocked},
        \ket{\pair\pair\blocked\fs\unoccupied\unoccupied\blocked},
        \ket{\pair\blocked\pair\fs\unoccupied\unoccupied\blocked},
        \ket{\blocked\pair\pair\fs\unoccupied\unoccupied\blocked},\\
        \ket{\pair\pair\unoccupied\fs\blocked\blocked\unoccupied},
        \ket{\pair\pair\blocked\fs\unoccupied\blocked\unoccupied},
        \ket{\pair\blocked\pair\fs\unoccupied\blocked\unoccupied},
        \ket{\blocked\pair\pair\fs\unoccupied\blocked\unoccupied},
        \ket{\pair\pair\blocked\fs\blocked\unoccupied\unoccupied},\\
        \ket{\pair\blocked\pair\fs\blocked\unoccupied\unoccupied},
        \ket{\blocked\pair\pair\fs\blocked\unoccupied\unoccupied},
        \ket{\pair\blocked\blocked\fs\pair\unoccupied\unoccupied},
        \ket{\blocked\pair\blocked\fs\pair\unoccupied\unoccupied},
        \ket{\blocked\blocked\pair\fs\pair\unoccupied\unoccupied}
      \},
    \end{split}
  \end{align}
  where {\blocked} denotes a blocked level.
  Blocked levels are occupied by unpaired electrons and thus have spin $1/2$.
  Thus, the true set of first-excited states includes a fourfold degeneracy for each of
  the states with two blocked levels.

  \subsection{Odd-parity sector}
  In the odd-parity sector there is always an odd number of blocked levels; thus, $k^{\rm
  eff}$ is of the form $n+1/2$ for $n\in \mathbb Z$.
  The subspace containing the ground state has $k^{\rm eff} = 1/2$.
  States in this subspace are obtained by blocking a single level at a time, and filling
  the lowest $\lfloor n_G/2\rfloor$ unblocked levels with pairs.
  This gives $N$ configurations, or $2N$ states including spin degeneracy.
  For $N=6$ and $n_G = 7$ we get

  \begin{align}
    \mathcal H^{\rm odd}_{k^{\rm eff}=1/2, g=0} = \{
      \ket{\pair\pair\pair\fs\unoccupied\unoccupied\blocked},
      \ket{\pair\pair\pair\fs\unoccupied\blocked\unoccupied},
      \ket{\pair\pair\pair\fs\blocked\unoccupied\unoccupied},
      \ket{\pair\pair\blocked\fs\pair\unoccupied\unoccupied},
      \ket{\pair\blocked\pair\fs\pair\unoccupied\unoccupied},
      \ket{\blocked\pair\pair\fs\pair\unoccupied\unoccupied}
    \} .
  \end{align}

  The $k^{\rm eff} = 3/2$ subspace is obtained by taking ${\mathcal H}_{k^{\rm eff} = 1,
  g = 0}^{\rm even}$ and adding a single blocked level at all possible level indices and
  shifting pairs to the lowest unoccupied level if necessary.

  \subsection{Degeneracy (grain only)}
  At fixed particle number, the number of states with $N$ levels and $k^{\rm eff} = k +
  \vert \B \vert / 2$ is equal to:
  \begin{align}
    D(N, k^{\rm eff}) &= \sum_{m=0}^{\lfloor k^{\rm eff} \rfloor} \ydim (N-2[k^{\rm
    eff} - m], m) \times p(N, 2[k^{\rm eff} - m]), \qquad p(N, q) = 2^q \binom{N}{q}\\
    &= \frac{N+1 - 2k^{\rm eff}}{N+1-k^{\rm eff}}\binom{2 N+1}{2 k^{\rm eff}}.
  \end{align}
  This expression does not depend on the number of electrons in the state, within
  the admissible domain $0 \leq 2k^{\rm eff} \leq \min(n_G, 2N-n_G)$ and $2k^{\rm eff}
  \equiv n_G \pmod 2$; it does not count all charge sectors at once.
  This expression is obtained by using the expression for $\ydim(N, q)$ in
  Ref.~\cite{roman2003}, which we introduced in Eq.~\eqref{eq:app:dNq}, and including the
  degeneracy due to the allowed number of blocked levels [accounted for in $p(N,q)$].

  Using the same examples as in the previous section, we find that $D(6, 1) = 65$, which
  is the cardinality of $\mathcal H^{\rm even}_{k^{\rm eff} = 1, g = 0}$, once the
  fourfold degeneracy of the blocked levels due to spin is taken into account.
  Additionally, $D(6, 1/2) = 12$, which is the cardinality of $\mathcal H^{\rm
  odd}_{k^{\rm eff}=1/2, g=0}$ (once the spin degeneracy is accounted for).

  \subsection{Degeneracy of the retained basis (including dot)}  \label{app:dim_ret}
  The degeneracy $D(N, k^{\rm eff})$ includes only the degeneracy for each $k^{\rm eff}$
  within the grain.
  The number of states in the $k^{\rm eff} \leq 1$ retained basis, however, additionally
  depends on the dot.

  We begin by assuming that $n_F$ is even.
  Defining $\mathcal H_{\rm ret.}$ as the retained basis, we can express its dimension in
  terms of $D(N, k^{\rm eff})$ and $D_{\rm dot}(k^{\rm eff})$:
  \begin{align}
    \dim \mathcal H_{\rm ret.} = \sum_{q \in \{0, 1/2, 1\}} D(N, q) D_{\rm dot}(q).
  \end{align}
  When there is an even number of electrons in the grain, there are also an even number of
  electrons in the dot.
  There are two possible dot states with even fermionic parity ($\ket0$ and $\ket2$);
  thus $D_{\rm dot}(0) = D_{\rm dot}(1) = 2$.
  Similarly, when there is an odd number of electrons in the grain, there must also be an
  odd number of electrons in the dot, of which there are again two possibilities
  ($\ket\uparrow$ and $\ket\downarrow$).
  Therefore, $D_{\rm dot}(1/2) = 2$, and
  \begin{align}
    \dim \mathcal H_{\rm ret.} = 2 \sum_{q \in \{0, 1/2, 1\}} D(N, q) = 2N(2N+1).
  \end{align}

  For $n_F$ odd, we obtain the same degeneracy; there are as many even-parity states as
  there are odd-parity states in the dot.

  Note that the evaluation of the sums above requires there to be available states in the
  grain; therefore, these results hold only at interior filling; when the grain is empty
  or full there will be fewer states in the retained basis.
  In the numerics relevant to this work, we are always near half filling, and thus
  $2N(2N+1)$ describes the sizes of our retained bases.

  \subsection{Verification}\label{app:convergence}

  In this section, we quantify the projection error by comparing the energies and
  current-flux relation with those obtained from exact diagonalization (ED).
  We use $g = 1.2d$ in this section of the Appendix, as in the main text.
  However, for such small systems (systems small enough for ED), $\Delta \sim d$ here,
  which implies that the $k = 1$ and $k = 2$ bands are not well separated.
  Therefore, this small-system analysis provides a concise benchmark of the projection
  error. For larger systems the retained basis is better separated from the remaining
  states in the Hilbert space, which is favorable for the projection, although the
  actual error also depends on the retained-to-discarded tunneling matrix elements.
  Furthermore, the errors here are associated with the projection procedure; the central
  result of this work, the determinant expression for fermionic matrix elements, remains exact.

  \subsubsection{Energies}
  For small systems ($N=8$ with $n_F = 7$ and $n_F = 8$), where the full spectrum is
  accessible by ED, we compare the projected ground- and excited-state energies with
  the exact results as a function of $\lambda/\Delta_{\rm ML}$.
  Here, $\Delta_{\rm ML}$ denotes the Matveev–Larkin parameter~\cite{matveev1997}, defined as
  \begin{align}
    \Delta_{\rm ML} = \frac{1}{2}\left(E_+ + E_-\right),
  \end{align}
  where $E_+$ ($E_-$) denotes the energy of the lowest-lying single-particle
  (single-hole) excitation relative to the ground state at half filling.
  $\Delta_{\rm ML}$ is the even-centered parity parameter. When the reservoir chemical
  potential is in the middle of the gap, the lowest electron-addition and electron-removal
  excitations lie exactly at $\Delta_{\rm ML}$.

  For sufficiently small $\lambda/\Delta_{\rm ML}$, the normalized energy error
  $\delta E/\Delta_{
  \rm ML}$ scales as $\mathcal O(\lambda^2/\Delta_{\rm ML}^2)$ with a small prefactor,
  as shown in Fig.~\ref{app:fig:verif}.
  At larger $\lambda/\Delta_{\rm ML}$, deviations from quadratic scaling become
  visible; the errors develop an approximately quartic dependence at the larger values
  of $\lambda/\Delta_{\rm ML}$ in the plotted range (up to $\simeq 0.5$).
  For the energy benchmarks shown here, and over the range of $\lambda / \Delta_{\rm
  ML}$ used in the main-text calculations, the normalized energy error remains well below
  $1\%$, whereas the current-flux relation exhibits larger errors.

  We note that for sufficiently large systems $\Delta \approx \Delta_{\rm ML}$;
  however, when the number of levels is on the order of unity and when $\Delta \sim d$
  it is more appropriate to use the Matveev--Larkin parameter, $\Delta_{\rm ML}$.

  \subsubsection{Energy-flux relation}

  Similarly, the normalized deviations in the energy-flux relation and in the
  current-flux relation are quantified in Fig.~\ref{app:fig:verif_flux}.
  The deviation in the energy-flux relation is taken relative to $\Delta_{\rm ML}$,
  whereas the deviation in the current-flux relation is taken relative to the ED
  critical current.

  \begin{figure}
    \centering
    \includegraphics[width=0.9\linewidth]{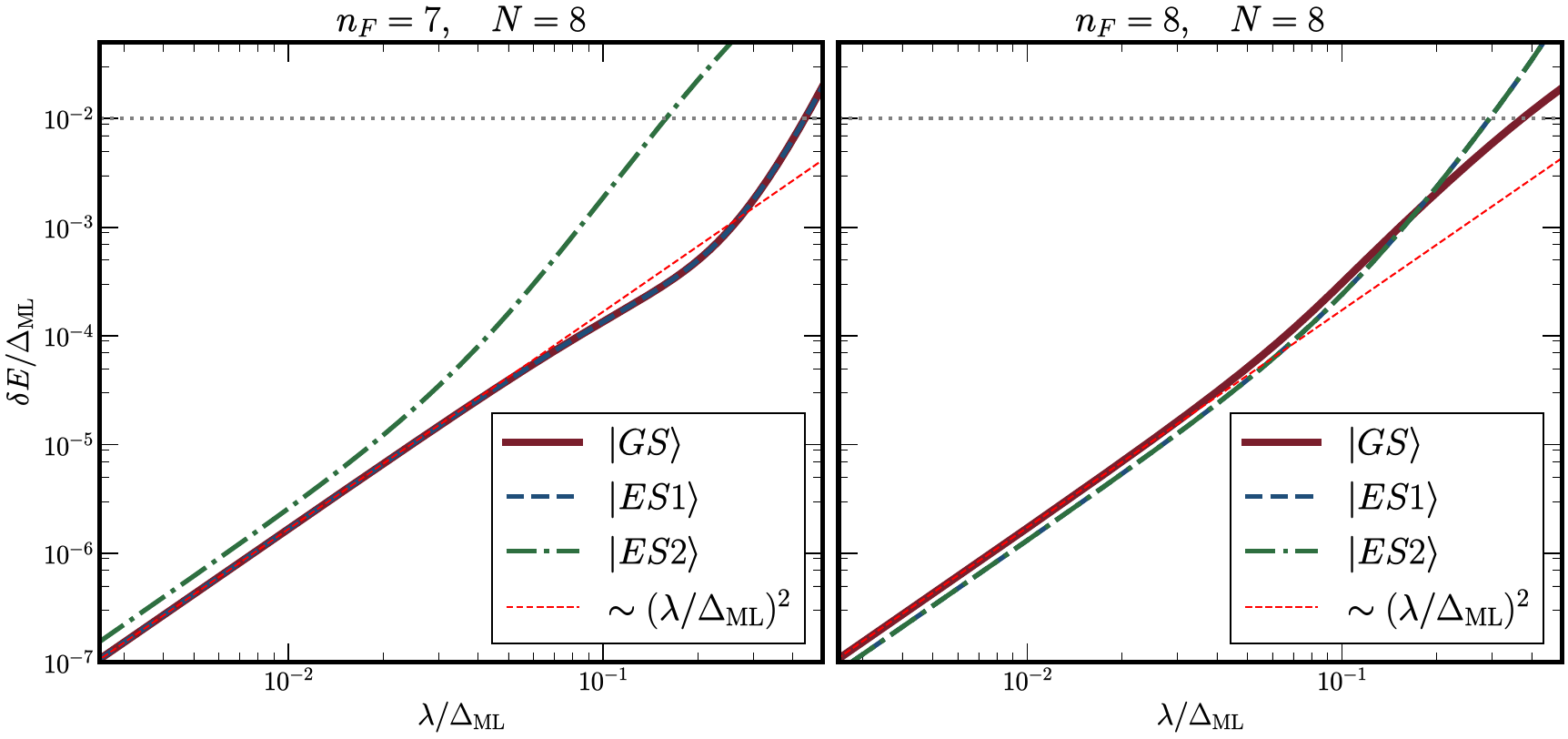}
    \caption{
      Projection error $\delta E/\Delta_{\rm ML}$ as a function of $\lambda/\Delta_{\rm
      ML}$ for $N=8$ levels with $n_F = 7$ (left) and $n_F=8$ (right), with $\delta
      E=\vert E_{\rm proj.}-E_{\rm ED}\vert$. The projected Hilbert space contains states
      satisfying $k_{\alpha}^{\rm eff}\leq 1$. Solid thick red curves show the
      ground-state error, while dashed thick curves show the errors of the first two
      excited states. The horizontal gray dotted line marks $1\%$; the red dashed lines
      are quadratic fits, $0.017(\lambda/\Delta_{\rm ML})^2$ in both cases (fitted for
      $\lambda/\Delta_{\rm ML} < 0.02$). Parameters are $U = 1.72\Delta_{\rm ML}$, $V =
      1.29\Delta_{\rm ML}$, $g = 1.2d$, $\Delta_{\rm ML} = 4.65d$, $n_{0,D} = 1$,
      $n_{0,G} = 6$ (left) and $n_{0,G} = 7$ (right).
    } \label{app:fig:verif}
  \end{figure}
  \FloatBarrier
  \begin{figure}
    \centering
    \includegraphics[width=0.8\linewidth]{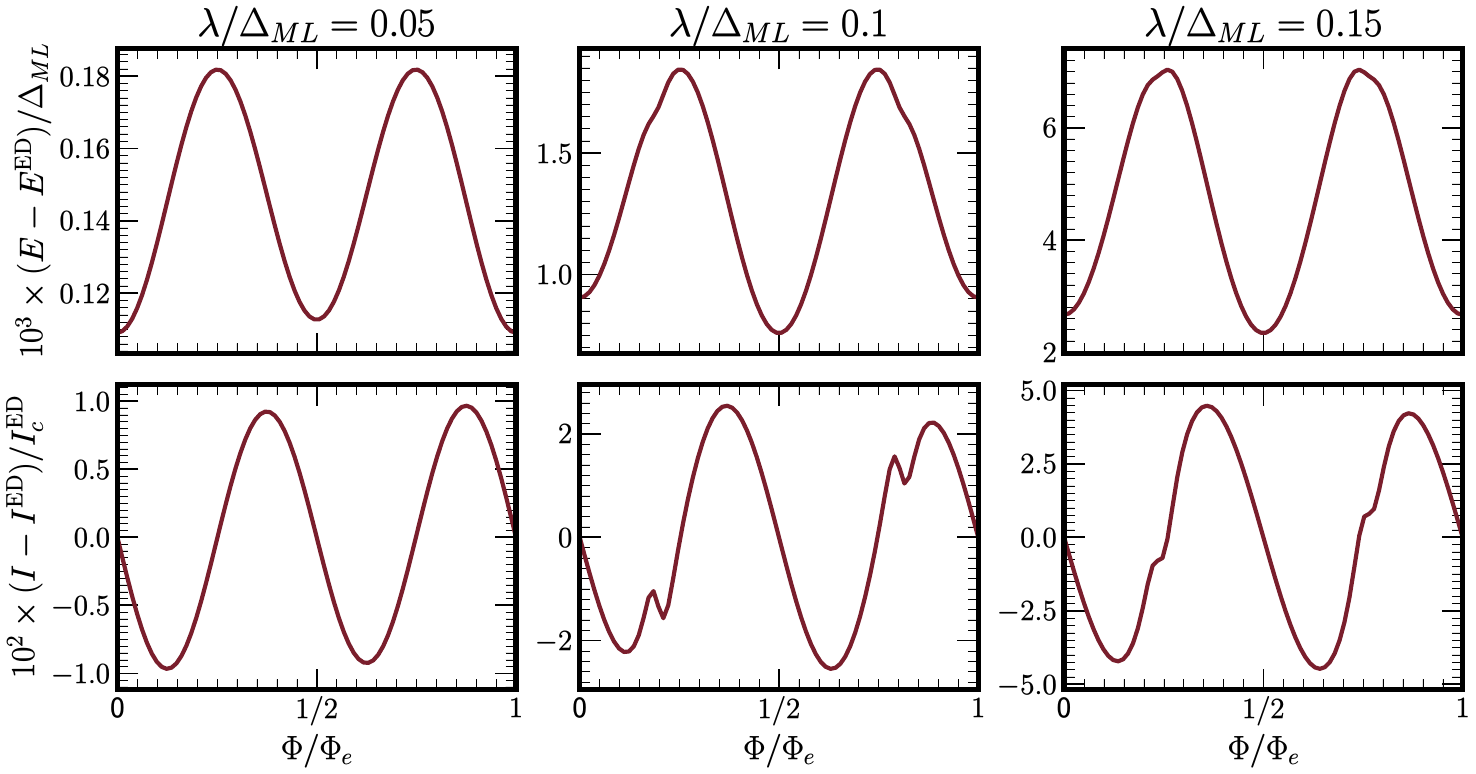}
    \caption{
      Projection error in the energy- and current-flux relations: the top row shows
      the energy error, relative to $\Delta_{\rm ML}$, and the bottom row shows the
      error in the current-flux relation, relative to the ED critical current $I^{\rm
      ED}_c$ (defined as the maximum absolute current $I^{\rm ED}_c = \max_{\Phi}
      \vert I^{\rm ED}(\Phi) \vert$). The projected Hilbert space contains states
      satisfying $k_{\alpha}^{\rm
      eff}\leq 1$. The ratio $\lambda/\Delta_{\rm ML}$ for each column is shown in the
      subfigure titles. Parameters are $U = 1.72\Delta_{\rm ML}$, $V = 1.29\Delta_{\rm
      ML}$, $g = 1.2d$, $\Delta_{\rm ML} = 4.65d$, $N = n_F = 8$, $n_{0, D} = 1$, and $n_{0,G} = 7$.
    }\label{app:fig:verif_flux}
  \end{figure}

  \subsection{Optical conductivity}\label{app:sec:optical_conductivity}

  In this section we give further information on the numerical calculation of the
  optical conductivity.
  A natural starting point is the Lehmann representation of the current-current
  correlation function, as in Eq.~\eqref{eq:pi}:
  \begin{align}
    \Pi_0^R(\omega) = \sum_{m \neq 0} \langle 0 \vert \curr_0 \left( \vert m \rangle
      \left[ \frac{1}{\hbar\omega + i \eta + \delta E_{m,0}} - \frac{1}{\hbar\omega + i
    \eta - \delta E_{m,0}} \right] \bra m \right ) \curr_0 \ket 0,
  \end{align}
  where $\eta$ is the artificial broadening that improves the visibility of the peaks,
  taken to be $\eta = \Delta/100$ in Fig.~\ref{fig:conductivity}.
  The operator in the parentheses can be written as $\hat{\mathcal G}_+(\omega) - \hat
  {\mathcal G}_-(\omega)$, where
  \begin{align}
    \hat {\mathcal G}_\pm(\omega) = \frac{1}{\hbar \omega + i \eta \pm (\hrga - E_0)},
  \end{align}
  where $\hrga$ is defined as in the main text [see Eq.~\eqref{eq:anderson}].
  Thus, by defining $\ket j = \curr_0 \ket 0$, the current-current correlation function
  can be rewritten as
  \begin{align}
    \Pi_0^R(\omega) = \bra j \ket{x_+(\omega)} - \bra j \ket{x_-(\omega)},
  \end{align}
  where
  \begin{align}
    \hat{\mathcal G}_\pm(\omega) \ket j = \ket{x_\pm(\omega)}.
  \end{align}
  This approach is advantageous because $\ket{x_\pm}$ can be calculated by solving a
  linear system that depends on otherwise known quantities:
  \begin{align}
    \left[\hbar \omega + i \eta \pm (\hrga - E_0) \right] \ket{x_\pm(\omega)} = \ket j.
  \end{align}
  This linear system is solved using SciPy~\cite{scipy}.
  Importantly, this approach allows for the current-current correlation function to be
  calculated without requiring the full spectrum, which makes it appropriate
  for sparse Hamiltonians.

  Thus, the broadened numerical form of the optical conductivity is
  \begin{align}
    \sigma(\omega) = \frac{\hbar}{i(\hbar\omega + i \eta)} \left[ \bra j
    \ket{x_+(\omega)} - \bra j \ket{x_-(\omega)} + \mean{\hat K_0} \right].
  \end{align}

  \section{Inner products \& matrix elements}
  \label{app:matrix_element_c}

  \subsection{Inner products}
  The expression cited in the main text [Eq.~\eqref{eq:innerproduct}] is a small
  generalization of the results from Refs.~\cite{faribault2012, claeys2015a,
  claeys2017, johnson2025}.
  In particular, we include blocked levels in this work; this section is intended to
  explicitly prove the statement that the inner product is zero unless $\C_\alpha = \C_\beta$.
  The inner product is zero for $M_\alpha \neq M_\beta$; therefore, we take $M \equiv
  M_\alpha = M_\beta$ throughout this section.
  We begin with the general expression for the inner product:
  \begin{align}
    & \bra{n_{G, \alpha}, k_\alpha, a_{\alpha}; \B_\alpha} \left( \prod_{(n, \sigma_n)
    \in \C_\alpha} \cop^\dagger_{n \sigma_n}\right)^\dagger \prod_{(m, \sigma_m) \in
    \C_\beta} \cop^\dagger_{m \sigma_m} \ket{n_{G, \beta}, k_\beta, a_{\beta}; \B_\beta} \\
    &= A_\alpha A_\beta \bra\vac\prod_{\nu = 1}^M \left[ \frac g2 \sum_{n\in \nB_\alpha}
    \frac{\hat S_n^-}{\epsilon_n - u_\nu^\alpha}\right]\left( \prod_{(n, \sigma_n) \in
    \C_\alpha} \cop^\dagger_{n \sigma_n}\right)^\dagger \prod_{(m,\sigma_m) \in \C_\beta}
    \cop^\dagger_{m \sigma_m} \prod_{\mu = 1}^M \left[ \frac g2\sum_{m \in \nB_\beta}
    \frac{\hat S_m^+}{\epsilon_m - u_\mu^\beta}\right] \ket \vac.
  \end{align}
  Consider a single annihilation operator $\cop_{\kappa, \sigma_\kappa}$ that annihilates
  an unpaired electron in $\bra{\Psi_\alpha}$ [thus $(\kappa, \sigma_\kappa) \in \C_\alpha$]:
  \begin{itemize}
    \item If $(\kappa, \sigma_\kappa) \in \C_\beta$, anticommuting $\cop_{\kappa,
      \sigma_\kappa}$ from $\bra{\Psi_\alpha}$ past $\cop^\dagger_{\kappa,
      \sigma_\kappa}$ in $\ket{\Psi_\beta}$ produces a term of the form $1 -
      \cop^\dagger_{\kappa, \sigma_\kappa} \cop_{\kappa, \sigma_\kappa}$.
      However, $\cop_{\kappa,\sigma_\kappa}$ can be anticommuted all the way to
      $\ket\vac$ (because $\kappa$ is not present in $\ket{n_{G, \beta}, k_\beta,
      a_{\beta}; \B_\beta}$, by the assumption that $\kappa \notin \nB_\beta$).
      Therefore, if $\kappa \in \B_\alpha$ and $\kappa \in \B_\beta$, the level $\kappa$
      only contributes through the scalar part ($= 1$) that arises when (anti)commuting
      $\cop_{\kappa, \sigma_\kappa}$ with $\cop^\dagger_{\kappa, \sigma_\kappa}$.
    \item On the other hand, if $(\kappa, \sigma_\kappa) \in \C_\alpha$ but $\kappa  \in
      \nB_\beta$, we will be able to commute $\cop_{\kappa, \sigma_\kappa}$ through the
      blocked levels of $\ket{\Psi_\beta}$. However, by assumption, $\kappa \in
      \nB_\beta$; therefore, we need to commute it with the pair creation operators:
      \begin{align}
        \cop_{\kappa, \sigma_\kappa}\left[\sum_{m\in \nB_\beta} \frac{\hat
        S_m^+}{\epsilon_m - u_\nu^\beta}\right] = \sigma_{\kappa}
        \frac{\cop^\dagger_{\kappa, \bar \sigma_\kappa}}{\epsilon_\kappa - u^\beta_\nu} +
        \left[\sum_{m\in \nB_\beta} \frac{\hat S_m^+}{\epsilon_m - u_\nu^\beta}\right]
        \cop_{\kappa, \sigma_\kappa}.
      \end{align}
      The term $\propto\cop_{\kappa, \bar\sigma_\kappa}^\dagger$ can be anticommuted all
      the way back to $\bra\vac$, annihilating it, whereas the second term will generate
      terms proportional to $\cop_{\kappa, \sigma_\kappa}$ (all of which annihilate $\ket\vac$).
      For the terms in $\ket{n_{G, \beta}, k_\beta, a_{\beta}; \B_\beta}$ not
      containing $\hat S^+_\kappa$, $\cop_{\kappa,\sigma_\kappa}$ annihilates $\ket\vac$.
    \item If $(\kappa, \sigma_\kappa) \in \C_\alpha$ and $(\kappa, \bar\sigma_\kappa) \in
      \C_\beta$, the blocked-level fermionic operators at level $\kappa$ can be commuted
      without a contraction term and thus annihilate the vacuum.
  \end{itemize}
  For a nonzero overlap, therefore, $\C_\alpha = \C_\beta$.
  Furthermore, only the vacuum expectation value of $\left( \prod_{(n,\sigma_n) \in
  \C_\alpha} \cop^\dagger_{n \sigma_n}\right)^\dagger \prod_{(m,\sigma_m) \in \C_\beta}
  \cop^\dagger_{m \sigma_m}$ contributes to the inner product.

  \subsection{Matrix elements}
  In order to derive the expression for the matrix elements used in the main text, we
  begin with the ansatz for the wave function, from Eq.~\eqref{eq:bethe_ansatz}:
  \begin{align}
    \ket{\Psi_{\alpha}} = A_\alpha\prod_{(b,\sigma_b) \in \C_{\alpha}}
    \cop_{b\sigma_b}^\dagger \left[ \prod_{\nu = 1}^{M_\alpha}{\frac g2\sum_{m \in
    \nB_{\alpha}}}\; \frac{\hat S_m^+}{\epsilon_m - u_\nu^{\alpha}} \right] \ket\vac.
  \end{align}
  Consider the matrix element of $\cop_{\kappa\sigma_\kappa}^\dagger$ between
  $\ket{\Psi_\alpha}$ and $\ket{\Psi_\beta}$ (we are considering $\bra{\Psi_\alpha}
  \cop_{\kappa\sigma_\kappa}^\dagger \ket{\Psi_\beta}$).
  This can be simplified immediately by recognizing that the matrix element is nonzero
  only if all creation operators in $\cop_{\kappa \sigma_\kappa}^\dagger
  \ket{\Psi_\beta}$ are contracted with the annihilation operators in $\bra{\Psi_\alpha}$.
  This can be satisfied in either of two cases:
  \begin{itemize}
    \item[A:] $\C_\alpha = \C_{\beta} \cup \{(\kappa, \sigma_\kappa)\}$ --- in this case,
      $\cop^\dagger_{\kappa \sigma_\kappa}$ removes a blocked level from
      $\bra{\Psi_\alpha}$, and the matrix element of $\cop^\dagger_{\kappa
      \sigma_\kappa}$ is related to an inner product of Bethe-ansatz wave functions;
      \begin{itemize}
        \item Here, $M_\alpha = M_\beta$.
      \end{itemize}
    \item[B:] $\C_\beta = \C_{\alpha} \cup \{ (\kappa, \bar \sigma_\kappa) \}$ and
      $\sigma_\kappa$ is opposite to the spin of the blocked level in $\ket{\Psi_\beta}$
      --- in this case, $\cop^\dagger_{\kappa \sigma_\kappa}$, together with the unpaired
      electron occupying level $\kappa$ in $\ket{\Psi_\beta}$, forms a pair and the
      matrix element of the creation operator is related to the matrix element of $\hat
      S^+_\kappa$.
      \begin{itemize}
        \item Here, $M_\alpha = M_\beta + 1$.
      \end{itemize}
  \end{itemize}
  In both cases, the matrix element will consist of two parts: (1) a sign factor $\Sigma$
  accounting for the normal ordering of the blocked-level creation/annihilation operators
  and (2) a (modified) $J$-determinant expression.
  Furthermore, the matrix element is zero unless $n_{G, \alpha} = n_{G, \beta} + 1$.

  \subsubsection{Condition A}
  The sign factor is simply given by
  \begin{align}
    \Sigma^{\kappa \sigma_\kappa}_{\alpha\beta} = \bra\vac \left(\prod_{(b, \sigma) \in
    \C_\alpha} \cop^\dagger_{b, \sigma}\right)^\dagger \cop_{\kappa,
    \sigma_\kappa}^\dagger \left(\prod_{(b',\sigma') \in \C_\beta} \cop_{b',
    \sigma'}^\dagger\right) \ket{\vac},
  \end{align}
  which follows directly from the expression for the wave functions and the fact that
  $\C_\alpha = \C_\beta \cup \{(\kappa, \sigma_\kappa)\}$.
  The modified $J$-determinant expression arises from the Bethe-ansatz part of the wave functions.
  Due to the operator $\cop_{\kappa, \sigma_\kappa}^\dagger$, the terms containing $\hat
  S_\kappa^+$ in the Bethe-ansatz part of $\ket{\Psi_\beta}$ cannot contribute to the
  matrix element.
  Thus, the $J$-determinant is simply obtained from
  \begin{align}
    \bra{n_{G, \alpha}, k_\alpha, a_{\alpha}; \B_\alpha} \ket{n_{G, \alpha}, k_\beta,
    a_{\beta}; \B_\beta\cup\{\kappa\}} = A_\alpha A_\beta \bra\vac\prod_{\nu = 1}^M
    \left[ \frac g2 \sum_{n\in \nB_\alpha} \frac{\hat S_n^-}{\epsilon_n -
    u_\nu^\alpha}\right] \prod_{\mu = 1}^M \left[ \frac g2 \sum_{m \in \nB_\beta
    \setminus \{ \kappa \}} \frac{\hat S_m^+}{\epsilon_m - u_\mu^\beta}\right] \ket \vac,
  \end{align}
  where, by convention, the number label of the $\beta$-side ket includes the blocked
  electron created by $\cop_{\kappa, \sigma_\kappa}^\dagger$, so both sides carry $n_{G,
  \alpha} = n_{G, \beta} + 1$ electrons; the restricted pair product itself is off shell
  and the on-shell labels $k_\beta, a_\beta$ refer to its parent state.
  However, $\ket{n_{G, \alpha},k_\alpha,a_{\alpha}; \B_\alpha}$ remains on shell,
  which is required by the inner product determinant expression we use below.
  In this context, `on shell' and `off shell' indicate whether the rapidities satisfy the
  Richardson equations (with the given set of blocked levels).

  Applying the expressions derived in Ref.~\cite{claeys2017} yields
  \begin{align}
    \bra{\Psi_\alpha} \cop_{\kappa, \sigma_\kappa}^\dagger \ket{\Psi_\beta} =
    (-1)^{N_\alpha^{\rm eff} + M_\alpha}\Sigma^{\kappa \sigma_\kappa}_{\alpha\beta}
    A_\alpha A_\beta\det J_{\alpha\beta}^\kappa,
  \end{align}
  where for $m,n \in \nB$, with $\nB = \nB_\alpha \cap \nB_\beta = \nB_\alpha$:
  \begin{align}
    (J_{\alpha \beta}^\kappa)_{mn} =
    \begin{cases}
      \ebv_m^\alpha + \ebv_m^\beta - 1 + \frac g2 \sum_{\ell \in \nB \setminus \{m\}}
      Z_{\ell m} & \text{ for } m = n; \\
      \frac g2 Z_{mn} & \text{ for } m \neq n.
    \end{cases}
  \end{align}

  The $J$-matrices depend only on $g$, $\epsilon_m$ and $\ebv_m$, all of which are real;
  therefore, in our chosen gauge $\bra{\Psi_\alpha} \cop_{\kappa, \sigma_\kappa}^\dagger
  \ket{\Psi_\beta} \in \mathbb R$.

  \subsubsection{Condition B}
  In this case, the sign factor is given by
  \begin{align}
    \begin{split}
      {\Sigma^{\kappa \sigma_\kappa}_{\alpha\beta}}' &= \bra\vac \left(\prod_{(b,\sigma)
      \in \C_\alpha} \cop^\dagger_{b \sigma}\right)^\dagger \hat S^-_\kappa
      \cop^\dagger_{\kappa\sigma_\kappa} \left(\prod_{(b',\sigma') \in \C_\beta} \cop_{b'
      \sigma'}^\dagger\right) \ket{\vac}\\
      &= \sigma_\kappa \bra\vac \left(\prod_{(b,\sigma) \in \C_\alpha} \cop^\dagger_{b
      \sigma}\right)^\dagger \cop_{\kappa\bar\sigma_\kappa} \left(\prod_{(b',\sigma') \in
      \C_\beta} \cop_{b' \sigma'}^\dagger\right) \ket{\vac}\\
      &= \sigma_\kappa \Sigma^{\kappa \bar \sigma_\kappa}_{\beta \alpha},
    \end{split}
  \end{align}
  where the final equality follows from $\Sigma^{\kappa \sigma_\kappa}_{\alpha \beta} \in
  \mathbb R$.

  After $\cop^\dagger_{\kappa\sigma_\kappa}$ creates the second electron at the blocked
  level $\kappa$, that level belongs to the paired subspace, within which $\{ \hat
  S^-_\kappa, \hat S^+_\kappa \} = 1$. Inserting this identity separates the matrix
  element into a blocked-level vacuum expectation value containing $\hat S^-_\kappa$, and
  a Bethe-ansatz matrix element containing $\hat S^+_\kappa$. Because $\hat S^+_\kappa$
  has even fermion parity, it commutes with the blocked-level operators. As before, the
  blocked-level vacuum expectation value is nonzero only when the remaining creation and
  annihilation operators can be contracted, thus enforcing condition B.

  The Bethe-ansatz part is given by
  \begin{align}
    \bra{n_{G, \alpha}, k_\alpha, a_{\alpha}; \B_\alpha} \hat S^+_{\kappa} \ket{n_{G,
    \beta}, k_\beta, a_{\beta}; \B_\beta} = A_\alpha A_\beta\bra\vac\prod_{\nu =
    1}^{M_\alpha} \left[ \frac g2 \sum_{n\in \nB_\alpha} \frac{\hat S_n^-}{\epsilon_n -
    u_\nu^\alpha}\right] \hat S^+_{\kappa} \prod_{\mu = 1}^{M_\beta} \left[ \frac g2
    \sum_{m \in \nB_\beta} \frac{\hat S_m^+}{\epsilon_m - u_\mu^\beta}\right] \ket \vac.
  \end{align}
  As discussed in Refs.~\cite{faribault2012, johnson2025}, this can be calculated
  by adding an auxiliary rapidity to $\ket{n_{G, \beta}, k_\beta, a_{\beta};
  \B_\beta}$, so that both states have the same number of pairs. Denoting the auxiliary
  rapidity by $w$, one obtains:
  \begin{align}
    \begin{split}
      \bra{n_{G, \alpha}, k_\alpha, a_{\alpha}; \B_\alpha} \hat S^+_{\kappa} & \ket{n_{G,
      \beta}, k_\beta, a_{\beta}; \B_\beta} \\
      & = A_\alpha A_\beta \lim_{w \rightarrow \epsilon_\kappa} (\epsilon_\kappa -
      w )\bra\vac\prod_{\nu = 1}^{M_\alpha} \left[ \frac g2 \sum_{n\in \nB_\alpha}
      \frac{\hat S_n^-}{\epsilon_n - u_\nu^\alpha}\right] \left[ \sum_{p
      \in\nB_\alpha} \frac{\hat S^+_{p}}{\epsilon_p - w} \right] \prod_{\mu =
      1}^{M_\beta} \left[ \frac g2 \sum_{m \in \nB_\alpha} \frac{\hat
      S_m^+}{\epsilon_m - u_\mu^\beta}\right] \ket \vac,
    \end{split}
  \end{align}
  where we have extended the $p$- and $m$-sums so that they include the level
  $\kappa$ and the $w \rightarrow \epsilon_\kappa$ residue gives us $\hat S^+_\kappa$;
  the added terms in the $m$-sum do not contribute to the residue, but are required for
  the validity of the determinant expression for the matrix element.
  Because the $\ket{n_{G, \alpha}, k_\alpha, a_{\alpha}; \B_\alpha}$ state is on shell,
  the determinant expression in Ref.~\cite{claeys2017} applies:
  \begin{align}
    \bra{n_{G, \alpha}, k_\alpha, a_{\alpha}; \B_\alpha} \hat S^+_{\kappa} \ket{n_{G,
    \beta}, k_\beta, a_{\beta}; \B_\beta} = \frac2g (-1)^{N^{\rm eff}_\alpha + M_\alpha}
    A_\alpha A_\beta \lim_{w \rightarrow \epsilon_\kappa} (\epsilon_\kappa -
    w ) \det \tilde J_{\alpha \beta}( w ).
  \end{align}
  This approach of expressing $\hat S^+_\kappa$ as a residue stems from
  Refs.~\cite{faribault2012, johnson2025}.
  The normalization does not depend on $w$, and the factor $2/g$ arises because the
  auxiliary pair operator is unscaled, whereas the determinant expression uses pair
  operators scaled by $g/2$.

  Here, for $m,n \in \nB_{\alpha}$:
  \begin{align}
    [\tilde J_{\alpha \beta}( w )]_{mn} =
    \begin{cases}
      \ebv_m^\alpha + \ebv_m^\beta + \frac g2 \frac{1}{\epsilon_m - w} - 1 + \frac g2
      \sum_{\ell \in \nB_\alpha \setminus \{m\} } Z_{\ell m} & \text{ for } m = n; \\
      \frac g2 Z_{mn} & \text{ for } m \neq n.
    \end{cases}
  \end{align}
  Thus,
  \begin{align}
    \frac2g \lim_{w \rightarrow \epsilon_\kappa} (\epsilon_\kappa - w ) \det
    \tilde J_{\alpha \beta}( w ) =
    \det J_{\alpha \beta}^\kappa.
  \end{align}
  Importantly, the $\epsilon_\kappa$-dependent regular part of the residue (on the
  diagonal) cancels, leaving only $\det J_{\alpha\beta}^\kappa$.
  Thus, the Bethe-ansatz contributions are identical in conditions A and B --- the
  expressions differ only in the sign factors $\Sigma$ vs $\Sigma'$.

  \subsubsection{Note on the numerical implementation}
  In the numerical implementation, the signs for conditions A ($\Sigma$) and B ($\Sigma'$)
  are calculated using bit masks and Hamming weights that count the number of fermionic
  operators that need to be anticommuted.
  For condition B, the $\hat S^-_\kappa$ is treated by adding a pair at level $\kappa$ in
  state $\ket{\Psi_\alpha}$ at $g = 0$.
  Thus, the fermionic operators never need to be constructed explicitly.

  \section{Perturbation theory: single-electron motion}\label{app:sec:pert}

  Suppose that the energy-phase relation [see Fig.~\ref{fig:flux}(\fluxepr)] can be written as
  \begin{align}
    E(\phi) = E_1 \cos\phi + E_2 \cos2\phi, \qquad \phi = 2 \pi \frac{\Phi}{\Phi_e},
  \end{align}
  where $E_1,E_2 \in \mathbb R$ are Fourier coefficients. The $E_1\cos\phi$ term is
  $2\pi$-periodic, implying single-electron motion, whereas the $E_2\cos2\phi$ term is
  $\pi$-periodic, consistent with motion of two electrons at a time; it can receive
  contributions from pair transfer as well as higher-order single-electron processes. We
  subtract a flux-independent constant and neglect higher harmonics. Interior extrema
  satisfy $\cos\phi = -E_1/(4E_2)$: for $E_2>0$ interior minima exist when $|E_1| <
  4|E_2|$ and sit exactly at $\phi=\pm\pi/2$ only for $E_1=0$. The energy-phase relation
  in Fig.~\ref{fig:flux}(\fluxweak, \fluxstrong) has the following properties:
  \begin{enumerate}
    \item The sign of $E_1$ depends on $n_{0, D}$ in the even-parity sector; the
      numerical results do not establish whether the same holds in the odd-parity
      sector.\label{point:alpha}
    \item In the odd-parity sector, $E_2 > 0$, whereas in the even-parity sector
      $E_2 < 0$.
      \begin{itemize}
        \item If $E_2 > 0$ and $\vert E_1 \vert \ll \vert E_2 \vert$, the minimum is
          at $\Phi \approx {h}/(4e)$, whereas when $E_2<0$, the minimum lies at
          $\Phi = 0$ or $\Phi = h/(2e)$.
        \item The numerics thus suggest that $\vert E_1\vert \ll \vert E_2 \vert$
          and $E_2 > 0$ in the odd-parity sector, because the minimum lies very close to
          $\Phi = h/(4e)$.
      \end{itemize}
  \end{enumerate}
  The second-order endpoint calculation below addresses only the leading
  single-electron harmonic $E_1$; it does not derive the coefficient $E_2$ of the
  $\Phi_{\rm sc}$-periodic component. As discussed above, the sign of $E_2$ can be
  inferred from the numerically calculated energy--flux relations.

  As can be seen in Fig.~\ref{fig:flux}(\fluxdepiction), there are two tunneling
  amplitudes, $\lambda_1$ and $\lambda_2$, and the Peierls phase can be assigned to
  either tunneling path following a gauge transformation.
  Let $\lambda_m(\Phi) = \lambda_{1, m} + \lambda_{2, m}(\Phi)$ parametrize the total
  tunneling [as in Eq.~\eqref{eq:fluxtun}]:
  \begin{align}
    {\htun = - \sum_{m\sigma} \left[ \lambda_m(\Phi) \cop^\dagger_{m \sigma} \d_\sigma
    + \lambda_m^*(\Phi) \d_\sigma^\dagger \cop_{m \sigma} \right].}
  \end{align}

  To explain the dependence of $E_1$ on $n_{0, D}$, we employ second-order
  perturbation theory.
  Treating the tunneling as a perturbation, we write:
  \begin{align}
    \hat H = \hat H_0 + \htun,
  \end{align}
  where
  \begin{align}
    \begin{split}
      \hat H_0 &= \sum_{m \sigma} \epsilon_m \cop^\dagger_{m\sigma} \cop_{m\sigma} - g
      \sum_{m\ell}\cop^\dagger_{m\uparrow}\cop^\dagger_{m\downarrow}\cop_{\ell\downarrow}
      \cop_{\ell\uparrow} + U(\hat n_D - n_{0, D} )^2.
    \end{split}
  \end{align}
  We neglect the charging in the grain, as it is not necessary to understand how the sign
  of $E_1$ depends on $n_{0, D}$.

  We work in the time domain and evaluate the regularized time integral to obtain the
  state energy shift. Thus, the perturbation-theory generator is

  \begin{align}
    \hat S(t) = -\frac{i}{\hbar} \int_0^t \nd t' e^{- 0^+ t' / \hbar} \htun(t'),
    \quad \text{with} \quad \htun(t') = \exp\left( i \hat H_0 t' / \hbar\right) \htun
    \exp\left(-i\hat H_0t' / \hbar\right).
  \end{align}
  To second order, the effective Hamiltonian is given by
  \begin{align}
    \hat H_{\rm eff}^{(2)}(t) = \frac12 \hat P [\hat S, \htun]\hat P = - \frac i{2\hbar}
    \int_0^t \nd t' e^{- 0^+ t' / \hbar} \hat P [\htun(t'), \htun(0)] \hat P,
  \end{align}
  where $\hat P = \ket{GS} \bra{GS}$ is the ground-state projector.
  We assume that the perturbation is switched on at $t = 0$, and then evaluate the
  Hamiltonian at time $t$.
  We are only interested in equilibrium properties of the system; thus, we will later
  take the limit $t \rightarrow \infty$, removing any transients.

  The unperturbed Hamiltonian $\hat H_0$ decouples into a grain part and a dot part,
  which implies that we can separate the time dependence as follows:
  \begin{align}
    \begin{split}
      \htun(t) &= - \sum_{m\sigma} \left[\lambda_m(\Phi)
        \cop^\dagger_{m\sigma}(t)\d_\sigma(t) +\lambda_m^*(\Phi) \d_\sigma^\dagger(t)
      \cop_{m\sigma}(t) \right],\\
      \d(t) &= \exp(i Ut (\hat n_D - n_{0, D} )^2 /\hbar) \d\exp(-i Ut (\hat n_D -
      n_{0, D} )^2 / \hbar ) = \hat{\mathcal U}_d(t) \d\, \hat{\mathcal U}_d(-t), \\
      \cop(t) &= \exp(i \hrg t / \hbar) \cop\exp(-i \hrg t / \hbar) = \hat{\mathcal
      U}_c(t) \cop\, \hat{\mathcal U}_c(-t).
    \end{split}
  \end{align}
  Equivalently, the decoupling of $\hat H_0$ implies that $[\hat{\mathcal U}_c(t),
  \hat{\mathcal U}_d(t')] = 0$ for all $t,t'$.

  We assume that the ground state has a well-defined number of electrons in the dot, and
  that it additionally factorizes into a product state of grain and dot.
  Large charging is an assumption selecting these endpoint configurations as the
  unperturbed ground states; number eigenstates of the decoupled Hamiltonian exist for
  any $U$. Quantitatively, we require $U, \Delta_{G,\pm} > \lambda$,
  where $\Delta_{G, \pm}$ denote the particle-addition and particle-removal gaps in
  the grain, respectively.
  Products of grain and dot eigenstates form the eigenbasis of $\hat H_0$.

  Consider a system with an even number of electrons and gate offset $n_{0, D}$ close to
  $0$ or $2$. For sufficiently large $U$, the ground-state dot occupation is pinned to
  $\langle \hat n_D \rangle = 0$ or 2, respectively. Consequently, $\delta n = \langle
  \hat n_D\rangle - n_{0, D}$ is small. The dot excitations thus cost $\Delta_{D, \pm} =
  U(1 \pm 2 \delta n) \approx U$, whereas the grain excitations cost $\Delta_{G, \pm}$; for
  simplicity, we assume that the relevant grain addition and removal excitation energies
  lie at exactly $\Delta_{G, +}$ and $\Delta_{G, -}$, respectively.

  The terms in $[\htun(t), \htun(0)]$ that contribute to the unperturbed
  ground-state expectation value are:
  \begin{align}
    \begin{split}
      \langle [\htun(t), \htun(0)] \rangle = \sum_{mn\sigma\sigma'} \lambda_m(\Phi)
      \lambda_{n}^*(\Phi) \langle [&
        \cop^\dagger_{m\sigma}(t) \cop_{n\sigma'} \d_\sigma(t) \d^\dagger_{\sigma'}
        + \cop_{n\sigma'}(t) \cop^\dagger_{m\sigma} \d^\dagger_{\sigma'}(t) \d_\sigma \\
        &- \cop^\dagger_{m\sigma} \cop_{n\sigma'}(t) \d_\sigma \d^\dagger_{\sigma'}(t)
        - \cop_{n\sigma'} \cop^\dagger_{m\sigma}(t) \d^\dagger_{\sigma'} \d_\sigma(t)
      ] \rangle.
    \end{split}
  \end{align}
  We approximate the time dependence of ground-state expectation values in the
  following way:
  \begin{align}
    \begin{split}
      \langle \hat d^\dagger(t) \hat d \rangle \simeq e^{-i \Delta_{D, -} t / \hbar}
      \langle \hat d^\dagger \hat d \rangle, \qquad \langle \hat d \hat d^\dagger(t)
      \rangle \simeq e^{i \Delta_{D, +} t /\hbar} \langle \hat d \hat d^\dagger \rangle, \\
      \langle \hat d^\dagger \hat d(t) \rangle \simeq e^{i \Delta_{D, -} t /\hbar}
      \langle \hat d^\dagger \hat d \rangle, \qquad \langle \hat d(t) \hat d^\dagger
      \rangle \simeq e^{- i \Delta_{D, +} t / \hbar} \langle \hat d \hat d^\dagger \rangle,
    \end{split}
  \end{align}
  where $\Delta_{D,\pm} \approx U$ when $\vert \delta n\vert \ll 1$, as assumed near
  integer gate offsets. This gives
  \begin{align}
    \begin{split}
      \langle [\htun(t), \htun(0)] \rangle \\
      = \sum_{nm\sigma\sigma'} \lambda_n(\Phi) \lambda_{m}^*(\Phi) \bigg[ &
        e^{-i \Delta_{D, +} t / \hbar}\langle \cop^\dagger_{n\sigma}(t)
        \cop_{m\sigma'} \d_\sigma \d^\dagger_{\sigma'} \rangle
        +e^{-i \Delta_{D, -} t / \hbar}\langle \cop_{m\sigma'}(t) \cop^\dagger_{n\sigma}
        \d^\dagger_{\sigma'} \d_\sigma \rangle \\
        & -e^{i \Delta_{D, +} t / \hbar}\langle \cop^\dagger_{n\sigma}
        \cop_{m\sigma'}(t) \d_\sigma \d^\dagger_{\sigma'} \rangle
        -e^{i\Delta_{D,-} t / \hbar}\langle \cop_{m\sigma'} \cop^\dagger_{n\sigma}(t)
        \d^\dagger_{\sigma'} \d_\sigma \rangle
      \bigg].
    \end{split}
  \end{align}
  Similarly, we approximate the time dependence of the grain part of ground-state
  expectation values as
  \begin{align}
    \begin{split}
      \langle \cop^\dagger(t) \cop \rangle \simeq e^{-i \Delta_{G,
      -}t/\hbar}\langle \cop^\dagger \cop\rangle, \qquad \langle \cop \cop^\dagger (t)
      \rangle \simeq e^{i \Delta_{G, +}t/\hbar}\langle \cop \cop^\dagger\rangle, \\
      \langle \cop^\dagger\cop(t)\rangle \simeq e^{i \Delta_{G, -}t/\hbar}\langle
      \cop^\dagger\cop\rangle, \qquad\langle \cop(t)\cop^\dagger\rangle \simeq e^{-i
      \Delta_{G, +}t/\hbar}\langle \cop\cop^\dagger\rangle,
    \end{split}
  \end{align}
  giving us
  \begin{align}
    \begin{split}
      \langle [\htun(t), \htun(0)] \rangle \\
      = \sum_{nm\sigma\sigma'} \lambda_n(\Phi) \lambda_{m}^*(\Phi) \bigg[ &
        e^{-i(\Delta_{G, -} + \Delta_{D, +} ) t / \hbar}\langle
        \cop^\dagger_{n\sigma} \cop_{m\sigma'} \d_\sigma \d^\dagger_{\sigma'} \rangle
        +e^{-i(\Delta_{G, +} + \Delta_{D,-}) t / \hbar}\langle \cop_{m\sigma'}
        \cop^\dagger_{n\sigma} \d^\dagger_{\sigma'} \d_\sigma \rangle \\
        & - e^{i(\Delta_{G, -} + \Delta_{D, +}) t / \hbar}\langle
        \cop^\dagger_{n\sigma} \cop_{m\sigma'} \d_\sigma \d^\dagger_{\sigma'} \rangle
        - e^{i(\Delta_{G, +} + \Delta_{D,-}) t / \hbar}\langle \cop_{m\sigma'}
        \cop^\dagger_{n\sigma} \d^\dagger_{\sigma'} \d_\sigma \rangle
      \bigg].
    \end{split}
  \end{align}
  This can be written as
  \begin{align}
    \begin{split}
      \langle [\htun(t), \htun(0)] \rangle = - 2i \sum_{nm\sigma\sigma'}
      \lambda_n(\Phi) \lambda_{m}^*(\Phi) \left[ \sin([\Delta_{G, -} + \Delta_{D,
        +} ]t / \hbar) \langle \cop^\dagger_{n\sigma} \cop_{m\sigma'} \d_\sigma
        \d^\dagger_{\sigma'} \rangle \right. \\
        \left. + \sin([\Delta_{G,+} + \Delta_{D, -} ]t / \hbar) \langle
      \cop_{m\sigma'} \cop_{n\sigma}^\dagger \d_{\sigma'}^\dagger \d_{\sigma}\rangle \right].
    \end{split}
  \end{align}
  Note that
  \begin{align}
    \hbar^{-1}\lim_{\varepsilon \rightarrow 0^+} \int_{0}^\infty \nd t\;
    e^{-\varepsilon t / \hbar} \sin(Et / \hbar) = \frac 1E,
  \end{align}
  which accounts for the energy denominators in the expressions below.
  Thus, when the dot is unoccupied, the energy denominator is $\Delta_{G, -} + \Delta_{D, +}$.
  For a doubly occupied dot, the corresponding denominator is $\Delta_{G, +} + \Delta_{D, -}$.
  In general, both terms contribute.

  However, to determine precisely where the parity transition happens, we would need to perform
  quasidegenerate perturbation theory instead, because this would require us to
  study the grain in the odd-parity sector, where the ground state is not well separated
  from excited states.

  Thus,
  \begin{align}
    \begin{split}
      &E_{GS, n_{0, D} = 0}^{(2)} = - \sum_{m} \frac{\vert \lambda_m(\Phi) \vert^2
      \mean{\hat n_m}}{\Delta_{G, -} + \Delta_{D, +}} \\
      &E_{GS, n_{0, D} = 2}^{(2)} = - \sum_{m} \frac{\vert \lambda_m(\Phi) \vert^2
      (2 - \mean{\hat n_m})}{\Delta_{G, +} + \Delta_{D, -}},
    \end{split}
  \end{align}
  where $\hat n_m = \sum_{\sigma = \uparrow,\downarrow} \cop^\dagger_{m \sigma}
  \cop_{m \sigma}$.
  We note that
  \begin{align}
    \vert \lambda_m(\Phi) \vert^2 = 2\lambda^2 [1 + \chi_m \cos(\phi) ],
  \end{align}
  where $\chi_m = \psi_{1, m} \psi_{2, m} = \pm 1$.
  If $\sum_m \chi_m = 0$, the coefficient $E_1$ (first harmonic) in the expression
  for the ground-state energy up to second order in $\lambda$ will have opposite signs at
  $n_{0, D} \approx 0$ and $n_{0, D} \approx 2$.
  Strictly speaking, the two limits involve different grain particle numbers: at fixed
  total electron number $n_F$, the sectors with $\mean{\hat n_D}\simeq0$ and
  $\mean{\hat n_D}\simeq2$ have grain occupations $\mean{\hat n_G}\simeq n_F$ and
  $\mean{\hat n_G}\simeq n_F-2$, respectively.
  The sign-reversal argument therefore assumes that the grain is large enough that
  adding or removing two electrons does not change the sign of the weighted sum $\sum_n
  \chi_n \langle \hat n_n\rangle$.
  Furthermore, the argument is inconclusive if this weighted sum vanishes.

\end{widetext}

\end{document}